\documentclass[prd,aps,showpacs,floats,floatfix,superscriptaddress,twocolumn,nofootinbib]{revtex4}

\usepackage{psfrag}
\usepackage[dvips]{graphicx}
\usepackage{amsmath}
\usepackage{amsfonts}
\usepackage{color}
\usepackage{dcolumn}
\usepackage{bm}
\usepackage{amsmath}

\newcommand{\hide}[1]{{}}

\def\laq{\raise 0.4ex\hbox{$<$}\kern -0.8em\lower 0.62ex\hbox{$\sim$}}
\def\gaq{\raise 0.4ex\hbox{$>$}\kern -0.7em\lower 0.62ex\hbox{$\sim$}}
\newcommand{\beq}{\begin{equation}}
\newcommand{\eeq}{\end{equation}}
\newcommand{\bea}{\begin{eqnarray}}
\newcommand{\eea}{\end{eqnarray}}
\newcommand{\ba}{\begin{array}}
\newcommand{\ea}{\end{array}}

\begin{document}

\title{A data-analysis driven comparison of analytic and numerical coalescing 
binary waveforms: nonspinning case}

\author{Yi Pan} 

\author{Alessandra Buonanno}

\affiliation{Department of Physics, University of Maryland, College Park, MD 20742}

\author{John G. Baker}

\author{Joan Centrella}

\author{Bernard J. Kelly}

\affiliation{Gravitational Astrophysics Laboratory, NASA 
Goddard Space Flight Center, 8800 Greenbelt Rd., Greenbelt, MD 20771}

\author{Sean T. McWilliams}

\affiliation{Department of Physics, University of Maryland, College Park, MD 20742}

\author{Frans Pretorius}

\affiliation{Department of Physics, Princeton University, Princeton, NJ 08544}

\author{James R. van Meter}

\affiliation{Gravitational Astrophysics Laboratory, NASA 
Goddard Space Flight Center, 8800 Greenbelt Rd., Greenbelt, MD 20771}

\affiliation{Center for Space Science \& Technology, 
University of Maryland Baltimore
County, Physics Department, 1000 Hilltop Circle, Baltimore, MD 21250}

\begin{abstract}
We compare waveforms obtained by numerically evolving nonspinning binary black holes to 
post-Newtonian (PN) template families currently used in the search for gravitational waves by 
ground-based detectors. We find that the time-domain 3.5PN template family, which 
includes the inspiral phase, has fitting factors (FFs) 
$ \geq 0.96$ for binary systems with total mass $M = 10 \mbox{--} 20 M_\odot$. 
The time-domain 3.5PN effective-one-body template family,
which includes the inspiral, merger and ring-down phases, 
gives satisfactory signal-matching performance with FFs 
$ \geq 0.96$ for binary systems with total mass $M = 10 \mbox{--}120M_\odot$.
If we introduce a cutoff frequency properly adjusted to the final black-hole 
ring-down frequency, we find that the frequency-domain stationary-phase-approximated template family 
at 3.5PN order has FFs $ \geq 0.96$ for binary systems with total mass $M = 
10 \mbox{--} 20 M_\odot$. However, to obtain high matching performances for larger binary masses, 
we need to either extend this family to unphysical regions of the parameter 
space or introduce a 4PN order coefficient in the frequency-domain 
GW phase. Finally, we find that the phenomenological Buonanno-Chen-Vallisneri 
family has FFs $\geq 0.97$ with total mass $M=10 \mbox{--} 120M_\odot$.  
The main analyses use the noise spectral-density of LIGO, but several tests are extended 
to VIRGO and advanced LIGO noise-spectral densities. 
\end{abstract}
\date{\today}


\maketitle

\section{Introduction}
\label{sec:intro}

The search for gravitational-waves (GWs) from coalescing binary systems  
with laser interferometer GW detectors~\cite{LIGO,GEO,TAMA,VIRGO,LISA} is based on 
the matched-filtering technique, which requires accurate knowledge of the 
waveform of the incoming signal. 
In the last couple of years there have been several breakthroughs
in numerical relativity (NR)~\cite{FP0, UTB0, Godd0}, and now independent groups
are able to simulate the inspiral, merger and ring-down phases of generic black- hole (BH) 
merger scenarios, including different spin orientations and mass ratios~\cite{mergers}.
However, the high computational cost of running such simulations makes it
difficult to generate sufficiently long inspiral waveforms that
cover the parameter space of astrophysical interest.

References~\cite{BCP,Goddshort} found good agreement between 
analytic (based on the post-Newtonian (PN) expansion) and numerical 
waveforms emitted during the inspiral, merger and ring-down phases 
of equal-mass, nonspinning binary BHs.   
Notably, the best agreement is obtained with 3PN or 3.5PN adiabatic 
waveforms~\cite{35PNnospin} (henceforth denoted as Taylor PN waveforms) 
and 3.5PN effective-one-body (EOB) waveforms~\cite{BD1,BD2,BD3,DJS,TD,BCD,DIS98}. 
In addition to the inspiral phase the latter waveforms include
the merger and ring-down phases.
Those comparisons suggested that it should be possible to 
design hybrid numerical/analytic templates, or even purely  
analytic templates with the full numerics used to guide the
patching together of the inspiral and ring-down waveforms. 
This is an important avenue to template construction as
eventually thousands of waveform 
templates may be needed to extract the signal from 
the noise, an impossible demand for NR alone. 
Once available, those templates could be used by 
ground-based laser interferometer GW detectors, such as LIGO, VIRGO, GEO and TAMA, and 
in the future by the laser interferometer space 
antenna (LISA) for detecting GWs emitted by solar mass
 and supermassive binary BHs, 
respectively. 

This paper presents a first attempt at investigating the closeness 
of the template families currently used in GW inspiral searches
to waveforms generated by NR simulations. Based on this investigation, we shall propose 
adjustments to the templates so that they include merger and ring-down phases.
In contrast, Ref.~\cite{baumgarte_et_al} examined the use of numerical
waveforms in inspiral searches, and compared numerical waveforms
to the ring-down templates currently used in burst searches. Similar to the
methodology presented here, fitting factors (FFs) 
[see Eq.~(\ref{FF}) below] are used in Ref.~\cite{baumgarte_et_al} 
to quantify the accuracy of numerical 
waveforms for the purpose of detection, as well
as the overlap of burst templates with the waveforms. 
Reference~\cite{baumgarte_et_al} found that by computing FFs between numerical 
waveforms from different resolution simulations of a given event, 
one can recast the numerical error as a maximum FF that the numerical 
waveform can resolve. In other words, any other template or putative
signal convolved with the highest resolution numerical simulation 
that gives a FF equal to or larger than this maximum FF is, for the purpose
of detection, indistinguishable from the numerical waveform. 
We will explore this aspect of the problem briefly. 
The primary conclusions we will draw from the analysis do not crucially depend on the exactness of the 
numerical waveforms. What counts here is that the templates can capture the 
dominant spectral characteristics of the true waveform.

For our analysis we shall focus on two nonspinning 
equal-mass binary simulation waveforms which differ in length,
initial conditions, and the evolution codes used to compute them:
Cook-Pfeiffer quasi-equilibrium initial data built on Refs.~\cite{CTS,GGB1,GGB2,PKST,CP}
evolved with Pretorius' generalized harmonic code~\cite{FP0},
and Brandt-Br\"ugmann puncture data~\cite{Brandt97b} evolved
using the Goddard group's moving-puncture code \cite{Godd0}. 
We also consider two nonspinning unequal-mass binary simulations with 
mass ratios $m_2/m_1 = 1.5$  and $m_2/m_1 = 2$ produced by the Goddard group.

The paper is organized as follows. In Sec.~\ref{sec2} we discuss 
the phase differences between PN inspiraling templates. In Sec.~\ref{sec3} 
we build {\it hybrid} waveforms by stitching together PN and NR waveforms. 
We try to understand how many NR cycles are needed to 
obtain good agreement between NR and PN waveforms, to offer
a guide for how long PN waveforms can be used as accurate templates.
In Sec.~\ref{sec4} we compute the FFs between
several PN template families and NR waveforms. We first 
focus on low-mass binary systems with total mass $M= 10 \mbox{--} 30 
M_\odot$, then high-mass binary systems with total mass $M=30 
\mbox{--} 120 M_\odot$.  Finally, Sec.~\ref{sec5} contains our main conclusions. 
In Appendix~\ref{appA} we comment on how different representations of 
the energy-balance equations give GW frequencies closer to or farther 
from the NR ones.

\section{Phase differences in post-Newtonian inspiraling models}
\label{sec2}

Starting from Ref.~\cite{3min}, which pointed out the importance of predicting 
GW phasing with the highest possible accuracy when building 
GW templates, many subsequent studies~\cite{DIS98,BD2,DIS01,DISspan,bcv1,DIJS,SPA,BCD} 
(those references are restricted to the nonspinning case) focused on 
this issue and thoroughly tested the accuracy of those templates, 
proposing improved representations of them. These questions were motivated 
by the observation that comparable-mass binary systems 
with total mass higher than $30 M_\odot$ merge in-band 
with the highest signal-to-noise ratio (SNR) for LIGO detectors, 
It follows that the corresponding templates 
demand  an improved analysis. 

In the absence of NR results and under the 
urgency of providing templates to search for comparable-mass binary BHs, 
the analytic PN community pushed PN calculations to higher PN orders, 
notably 3.5PN order~\cite{35PNnospin}, and also proposed ways 
of resumming the PN expansion, either 
for conservative dynamics (the EOB approach~\cite{BD1,DJS,TD}), 
radiation-reaction effects (the Pad{\'e} resummation~\cite{DIS98}), or 
both~\cite{BD2,BCD}. Those results lead to several conclusions: (i) 3PN terms improve 
the comparison between analytic and (numerical) 
quasi-equilibrium predictions~\cite{CP,GGB1,DGG,ICO}; (ii) Taylor expanded and 
resummed PN predictions for equal-mass binary systems are much closer at 3.5PN order 
than at previous PN orders, indicating a convergence 
between the different schemes~\cite{bcv1,DIS01,BCD}; (iii) 
the two-body motion is quasi-circular until the end of a rather 
blurred plunge~\cite{BD2}, (iv) the transition to ring-down can 
be described by an extremely short merger phase~\cite{BD2,BCD}.  
Today, with the NR results we are in a position to sharpen the above conclusions, and
to start to assess the closeness of analytic templates to numerical waveforms.  

Henceforth, we restrict the analysis to the three time-domain 
physical template families which are closest to NR results~\cite{BCP,Goddshort}: 
the adiabatic Taylor PN model (Tpn)~[see, e.g., Eqs.~(1), (10), and (11)--(13) in 
Ref.~\cite{bcv2}] computed at 3PN and 3.5PN order, 
and the nonadiabatic EOB model (Epn)~ [see e.g., Eqs.~(3.41)--(3.44) 
in Ref.~\cite{BD2}] computed at 3.5 PN order. We shall denote 
our models as Tpn(n) and Epn(n), n being the PN order.
The Tpn model is obtained by solving a particular 
representation of the balance equation. In Appendix~\ref{appA} we 
briefly discuss how time-domain PN models based on different 
representations of the energy-balance equation would 
compare with NR results. 

The waveforms we use are always derived in the so-called 
{\it restricted approximation} which uses the amplitude 
at Newtonian order and the phase at the highest PN order available. 
They are computed by solving PN dynamical equations 
providing the instantaneous frequency $\omega(t)$ and phase 
$\phi(t)=\phi_0+\int_{t_0}^t\omega(t')dt'$, thus 
\beq
\label{quadstrain}
h(t)={\cal A}\,\omega(t)^{2/3}\cos[2\phi(t)]\,, 
\eeq 
where $t_0$ and $\phi_0$ are the initial time and phase, 
respectively, and ${\cal A}$ is a constant amplitude, 
irrelevant to our discussion. The inclusion of higher-order 
PN corrections to the amplitude can be rather important 
for certain unequal-mass binary systems,
and will be the subject of a future study. 

When measuring the differences between waveforms we weight them
by the power spectral-density (PSDs) of the detector,  
and compute the widely used {\it fitting factor} (FF) 
(i.e., the ambiguity function or normalized overlap), or equivalently the 
{\it mismatch} defined as 1-FF. 
Following the standard formalism of matched-filtering 
[see, e.g., Refs.~\cite{DA,DIS98,bcv1}], we define the FF as the overlap 
$\langle h_1(t),h_2(t)\rangle$ between the waveforms $h_1(t)$ and
$h_2(t)$: 
\bea
\langle h_1(t),h_2(t)\rangle\equiv4\,{\rm Re}
\int_0^\infty\frac{\tilde{h}_1(f)\tilde{h}^*_2(f)}{S_h(f)}df\,, 
\qquad\quad \nonumber\\
{\rm FF}\equiv\max_{t_0,\phi_0,\lambda^i}
\frac{\langle h_1,h_2(t_0, \phi_0, \lambda^i)\rangle}
{\sqrt{\langle h_1,h_1\rangle\langle 
h_2(t_0, \phi_0, \lambda^i),h_2(t_0, \phi_0, \lambda^i)\rangle}}\,, 
\nonumber\\
\label{FF}
\eea 
where $\tilde{h}_i(f)$ is the Fourier transform of $h_i(t)$, and
$S_h(f)$ is the detector's PSD. Thus, the FF is the normalized
overlap between a target waveform $h_1(t)$ and a set of template
waveforms $h_2(t_0,\phi_0,\lambda^i)$ maximized over the initial time
$t_0$, initial phase $\phi_0$, and other parameters $\lambda^i$. 
Sometimes we are interested in FFs that are optimized {\it only} over 
$t_0$ and $\phi_0$; we shall denote these as ${\rm FF}_0$.
For data-analysis purposes, the FF has more direct meaning than 
the phase evolution of the waveforms, since it takes into account 
the PSDs and is proportional to the
SNR of the filtered signal. Since the event rate is proportional to 
the cube of the SNR, and thus to the cube of the FF, a FF$=0.97$ 
corresponds to a loss of event rates of $\sim 10\%$. 
A template waveform is considered a satisfactory representation 
of the target waveform when the FF is larger than 0.97. 

When comparing two families of waveforms, the FF is optimized over the
initial phase of the template waveform, and we also need to specify
the initial phase of the target waveform. Since there is no preferred
initial phase of the target, two options are usually adopted: (i) the
initial phase {\it maximizes} the FF or (ii) it {\it minimizes} the FF. 
The resulting FFs are referred to as the {\it best} and {\it minimax} 
FFs~\cite{DISspan}. All FFs we present in this paper are minimax
FFs. Although the FF of two waveform families is generally asymmetric
under interchange of the template family~\cite{bcv1}, the best and the
minimax FF$_0$s are symmetric (see Appendix B of Ref.~\cite{DISspan}
for details).  Henceforth, when comparing two waveform families using
FF$_0$, we do not need to specify which family is the target.

We shall consider three interferometric GW detectors:
LIGO, advanced LIGO and VIRGO. The latter two have better
low-frequency sensitivity and broader bandwidth. For LIGO, we use the
analytic fit to the LIGO design PSD given in Ref.~\cite{DIS01}; for
advanced LIGO we use the broadband configuration PSD given in
Ref.~\cite{AdvLIGO}; for VIRGO we use the PSD given in Ref.~\cite{DIS01}. 

\begin{figure}
\begin{center}
\includegraphics[width=0.45\textwidth]{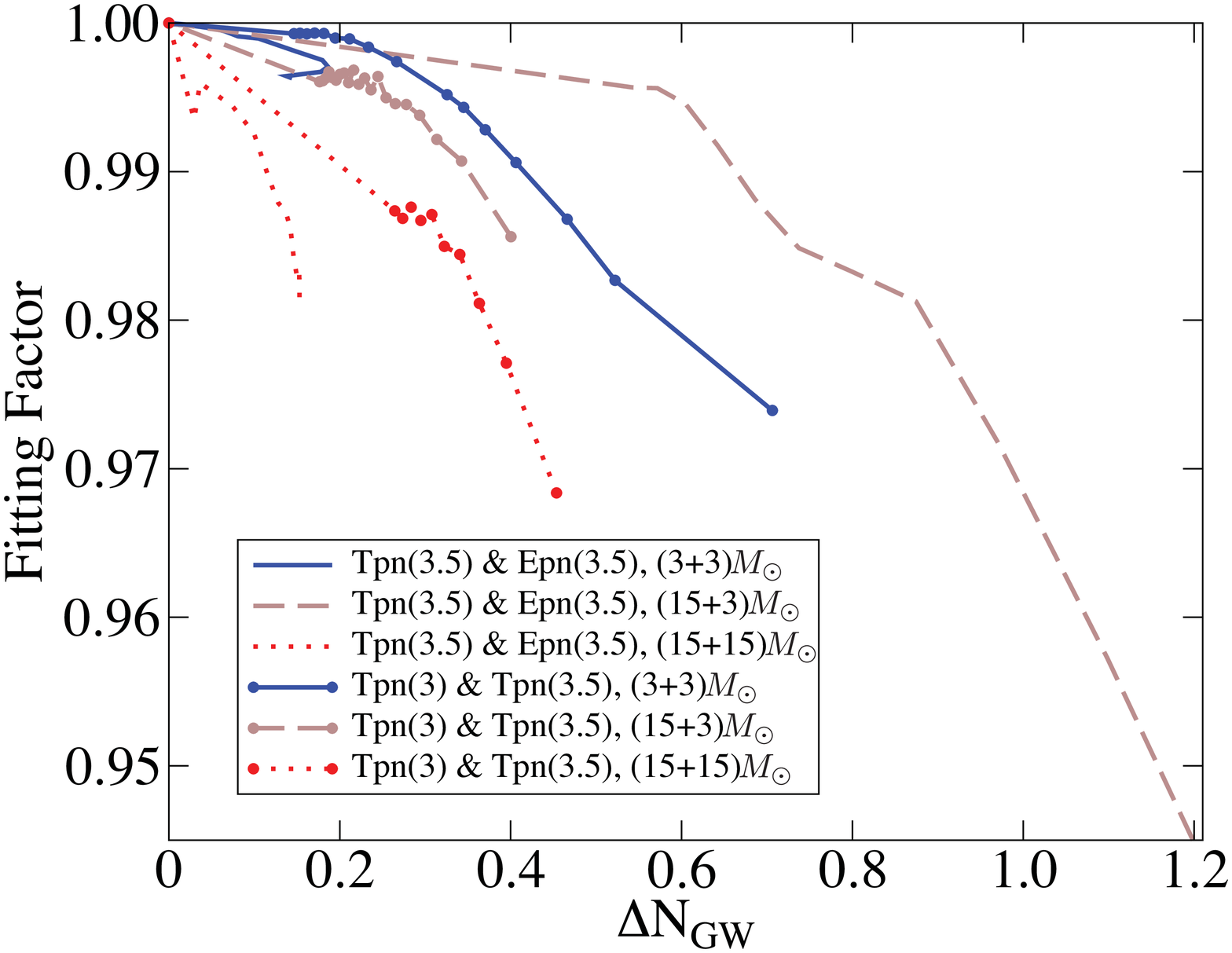}
\caption{
We show ${\rm FF}_0$s between waveforms generated from the three PN models Tpn(3), Tpn(3.5) and 
Epn(3.5) versus $\Delta {\rm N}_{\rm GW}$ [see Eq.~\eqref{dn}]. The ${\rm FF}_0$s are evaluated with LIGO's PSD. 
Note that for Tpn(3.5) and Epn(3.5) and a $(15+3)M_\odot$ binary, the lowest ${\rm FF}_0$ is 0.78 
and the difference in the number of GW cycles $\Delta {\rm N}_{\rm GW} \simeq 2$. In the 
limit $\Delta {\rm N}_{\rm GW}\rightarrow0$, the FF$_0$ goes to unity.
\label{fig:dphovp}}
\end{center}
\end{figure}

In Fig.~\ref{fig:dphovp}, we show the ${\rm FF}_0$s as functions of
the accumulated difference in the number of GW cycles between waveforms generated with
different inspiraling PN models and for binary systems with different component
masses. We first generate two waveforms by evolving two PN models, say, ``PN$_1$''
and ``PN$_2$'' which start at the same GW frequency $f_{\rm GW}=30$Hz and 
have the same initial phase. The two waveforms are terminated at the same 
ending frequency $f_{\rm GW}=f_{\rm end}$ up to a maximum $f_{\rm end,\rm max} 
= {\rm min}(f_{\rm end,{\rm PN}_1},f_{\rm end,{\rm PN}_2})$, 
where $f_{\rm end,{\rm PN}}$ is the frequency at which the PN inspiraling 
model ends. (For Tpn models this is the frequency at which the PN energy 
has a minimum; for Epn models it is the EOB light-ring frequency.)
Then, we compute the difference in phase and number of GW cycles 
accumulated until the ending frequency
\beq\label{dn}
\Delta {\rm N}_{\rm GW}= \frac{\Delta \phi}{\pi}=\frac{1}{\pi}\left[\phi_{\rm PN_1}(f_{\rm end})-
\phi_{\rm PN_2}(f_{\rm end})\right]\,. 
\eeq 
By varying $f_{\rm end}$ (up to $f_{\rm end,\rm max}$) $\Delta {\rm N}_{\rm GW}$ changes,
though not necessarily monotonically. 
Although there seems to be a loose correlation between the ${\rm FF}_0$s 
and $\Delta {\rm N}_{\rm GW}$, it is hard to quantify it as a one-to-one 
correspondence. For example, a phase difference of about
half a GW cycle ($\Delta {\rm N}_{\rm GW}\simeq0.5$) is usually thought to be
significant.  However, here we find relatively high ${\rm FF}_0$s
between $0.97$ and $>0.99$, depending on the masses of the
binary and the specific PN model used. This happens because the FF
between two waveforms is not determined by the total phase difference
accumulated, but rather by how the phase difference accumulates across
the detector's most sensitive frequency band. The relation between FFs
and phase differences is also blurred by the maximization over the
initial time and phase: shifting the phase by half a cycle from the
most sensitive band to a less sensitive band can increase the matching
significantly.
We conclude that with LIGO's PSD, 
after maximizing {\it only} on initial phase and time, 
Epn(3.5) and Tpn(3.5) templates are close to each other 
for comparable-mass binary systems $M = 6 \mbox{--} 30 M_\odot$ 
with FF$_0 \,\gaq \, 0.97$, but they can be different  
for mass ratios $m_2/m_1 \simeq 0.3$ with FF$_0$ 
as low as $\simeq 0.8$. 
Tpn(3) and Tpn(3.5) templates have FF$_0 \, \gaq\, 0.97$ for the 
binary masses considered. Note that for $m_2/m_1=1$ [$\simeq 0.3$] 
binary systems, Tpn(3.5) is closer to Epn(3.5) [Tpn(3)] 
than to Tpn(3) [Epn(3.5)]. Note also that 
when maximizing on binary masses the FFs can increase significantly, 
for instance, for a $(15+3)M_\odot$ binary, the FF between Tpn(3.5) and 
Epn(3.5) waveforms becomes $>0.995$, whereas FF$_0\simeq0.8$.

\section{Building and comparing hybrid waveforms}
\label{sec3} 

Recent comparisons~\cite{BCP, Goddshort} between analytic and 
numerical inspiraling waveforms of nonspinning, equal-mass binary systems have shown that numerical 
waveforms are in good agreement with Epn(3.5), Tpn(3) and Tpn(3.5) waveforms. 
Those results were assessed using eight and sixteen numerical inspiral GW cycles. Can we conclude 
from these analyses that Epn(3.5), Tpn(3.5) and Tpn(3) 
can safely be used to build a template bank for detecting {\it inspiraling} GW signals? 
A way to address this question is to compute the mismatch between hybrid waveforms 
built by attaching either Epn or Tpn waveforms to the {\em same} 
numerical waveform, and varying the time when the attachment is made. This
is equivalent to varying the number of numerical GW cycles $n$ in the hybrid template.
The larger $n$ the smaller the mismatch, as we are using the same numerical
segment in both waveforms. For a desired maximum mismatch, say $3\%$, we can
then find the smallest number $n$ of numerical cycles that is required
in the hybrid waveform. This number will, of course, depend on the binary mass
and the PSD of each detector.

\subsection{Hybrid waveforms}
\label{sec3.1}

We build hybrid waveforms by connecting PN waveforms to 
NR waveforms at a chosen point in the late inspiral stage.
As mentioned before, we use NR waveforms generated with 
Pretorius'~\cite{BCP} code and the Goddard group's \cite{Goddlong} code. 
Pretorius' waveform is from an 
equal-mass binary with total mass $M$, and equal, co-rotating spins 
($a=0.06$). The simulation lasts $\simeq 671M$, and the waveform has 
$\simeq 8$ cycles before the formation of the common apparent horizon. 
The Goddard waveform refers to an equal-mass nonspinning binary. 
The simulation lasts about $\simeq 1516M$, and the waveform 
has $\simeq 16$ cycles before merger.

\begin{figure}
\begin{center}
\includegraphics[width=0.45\textwidth]{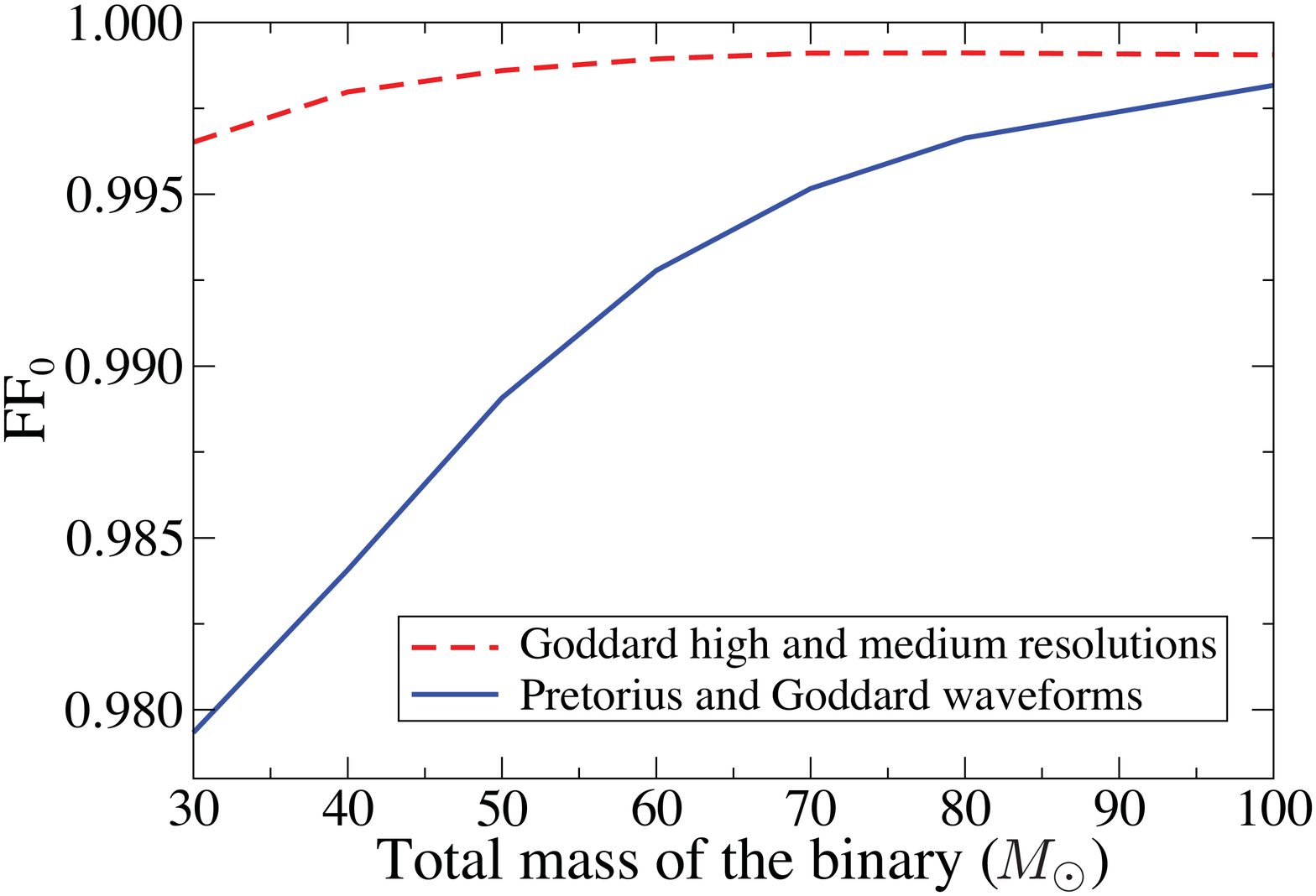}
\caption{${\rm FF}_0$ between NR waveforms as a function of the binary
  total-mass $M$. The solid curve are generated for waveforms from
  Pretorius and the Goddard group. 
  The longer Goddard waveform is
  shortened such that both waveforms last $\simeq 671M$ and contain
  $\simeq8$ cycles. The dashed curve is generated for waveforms from the
  high-resolution and medium resolution simulations of the Goddard
  group.  All ${\rm FF}$s are evaluated using LIGO's PSD.}
\label{fig:NRff}
\end{center}
\end{figure}

Since we will present results from these two waveforms it is useful 
to first compare them by computing the ${\rm FF}_0$. 
Although the binary parameters considered by Pretorius 
and Goddard are slightly different, we expect the waveforms, 
especially around the merger stage, to be fairly close. 
Comparisons between (shorter) waveforms computed with moving punctures 
and generalized-harmonic gauge were reported in Ref.~\cite{NRcomp}, 
where the authors discussed the different initial
conditions, wave extraction techniques, and compared the
phase, amplitude and frequency evolutions. 
Since the two simulations use different initial conditions and 
last for different amounts of time we cut the longer
Goddard waveform at roughly the frequency where the Pretorius waveform
starts. In this way we compare waveforms that have the same length 
between the initial time and the time at which the wave amplitude 
reaches its maximum. In Fig.~\ref{fig:NRff}, we show the ${\rm FF}_0$ as function 
of the total binary mass. Despite differences in the two simulations 
the ${\rm FF}_0$s are rather high. The waveforms differ more
significantly at lower frequencies. Indeed, as the total mass decreases
the ${\rm FF}_0$s also decrease as these early parts of the
waveform contribute more to the signal power given LIGO's PSD.

Any waveform extracted from a numerical simulation will inherit 
truncation errors, affecting both the waveform's 
amplitude and phase~\cite{BCP,Goddlong, baumgarte_et_al}. 
To check whether those differences would change the results 
of the comparisons between NR and PN waveforms, we plot 
in Fig.~\ref{fig:NRff} the ${\rm FF}_0$s versus total binary mass 
between two Goddard waveforms generated from a high and a medium resolution run~\cite{Goddlong}. 
The ${\rm FF}_0$s are extremely high ($>0.995$). 

Based on the comparisons between high and medium resolution waveforms,
we can estimate the FFs between high resolution and exact
waveforms. If we have several simulations with different resolutions,
specified by the mesh-spacings $x_i$, and $x_i$ are sufficiently
small, we can assume that the waveforms $h_i$ are given by
\beq 
h_i=h_0+x_i^n h_d\,, 
\eeq 
where $n$ is the convergence factor of
the waveform, $h_0$ is the exact waveform generated from the infinite
resolution run ($x_0\rightarrow0$), and $h_d$ is the leading 
order truncation error contribution to the waveform and is independent
of the mesh spacing $x_i$. We find that 
the mismatch between the waveforms $h_i$ and $h_j$, 
$1-{\rm FF}_{ij}$, scales as
\beq
1-{\rm FF}_{ij}\propto (x_i^n-x_j^n)^2\,. 
\eeq 
In the Goddard simulations, the high and medium resolution runs have mesh-spacing
ratio $x_h/x_m=5/6$, and the waveform convergence rate is $n=4$~\cite{Goddlong}. 
The FF between the high resolution and exact 
waveforms $h_h$ and $h_0$ is given by 
\beq 
{\rm FF}_{0h}=1-0.87 (1-{\rm FF}_{hm})\,, 
\eeq 
where FF$_{hm}$ is the FF between the high
and medium resolution waveforms $h_h$ and $h_m$. That is to say, the
mismatch between $h_h$ and $h_0$ is slightly smaller than that between
$h_h$ and $h_m$, where the latter can be derived from the FFs shown in 
Fig.~\ref{fig:NRff}.  Henceforth, we shall always use high-resolution
waveforms. A similar calculation for Pretorius' waveform gives
${\rm FF}_{0h}=1-0.64 (1-{\rm FF}_{hm})$, though here $x_h/x_m = 2/3$
and $n=2$. See Fig. 6 of Ref.~\cite{baumgarte_et_al} for a plot
of ${\rm FF}_{hm}$ calculated from the evolution of the Cook-Pfeiffer 
initial data~\footnote{The plot in Ref.~\cite{baumgarte_et_al} is for ``d=16''
corotating Cook-Pfeiffer initial data, whereas the results presented here
are from ``d=19'' initial data. However, the resolutions used for
both sets were the same, and thus the mismatches should be similar,
in particular in the higher mass range.}; there $FF_0$ ranges from
$\approx 0.97$ for $M/M_s=30$ to $\approx0.99$ for $M/M_s=100$. In other words,
the mismatch between Goddard's and Pretorius' waveform shown
in Fig.\ref{fig:NRff} is less than the estimated mismatch from
numerical error in the latter waveform. 

We build hybrid waveforms by stitching together the PN and NR waveforms computed 
for binary systems with the {\it same} parameters. At the point where we connect the two 
waveforms, we tune the initial time $t_0$ so that the frequency of the PN 
waveform is almost the same as the frequency of the NR waveform (there is a subtlety 
trying to match exactly the frequencies that is discussed at the end of this section). The initial phase $\phi_0$ 
is then chosen so that the strain of the hybrid waveform is continuous at 
the connecting point.

\begin{figure}
\begin{center}
\includegraphics[width=0.45\textwidth]{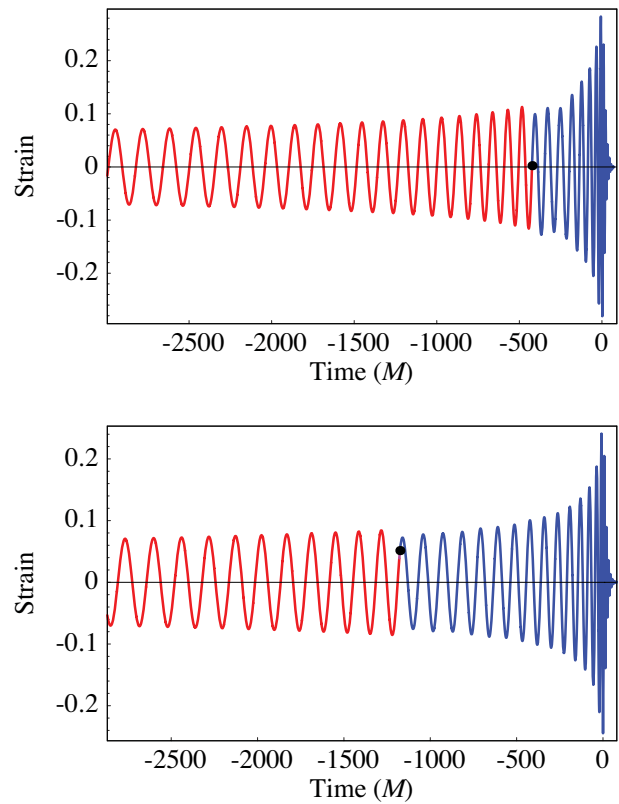}
\caption{We show two examples of hybrid waveforms, starting from 40Hz. The PN waveforms are generated 
with the Tpn(3.5) model, and the NR waveforms in the upper and lower panels are 
generated from Pretorius' and 
Goddard's simulations, respectively. We mark with a dot the point where we connect the PN and NR waveforms.
\label{fig:hybrid}}
\end{center}
\end{figure}

In Fig.~\ref{fig:hybrid}, we show two examples of hybrid waveforms of
an equal-mass binary. We stitch the waveforms at
points where effects due to the initial-data transient pulse are 
negligible. We find an amplitude difference on the order 
of $\sim10\%$ between the Goddard waveform  and 
the restricted PN waveform. This difference is also 
present in Pretorius' waveform, but it is somewhat compensated for by 
amplitude modulations caused by eccentricity in the initial data.
In Ref.~\cite{Goddlong} it was shown that PN waveforms 
with 2.5PN amplitude corrections give better agreement
(see e.g., Fig. 12 in Ref.~\cite{Goddlong}). 
However, the maximum amplitude errors in the waveforms 
are also on the order of 10\% \cite{BCP,Goddlong}. 
Since neither 2PN nor other lower PN order 
corrections to the amplitudes are closer to the 2.5PN order,
we cannot conclude that 2.5PN amplitude corrections 
best approximate the numerical waves. 
Thus, we decide to use 
two sets of hybrid waveforms: one constructed with 
restricted PN waveforms, and the other
with restricted PN waveforms rescaled by a single amplitude factor,
which eliminates amplitude differences with the NR waveforms. 
We shall see that the difference between these two cases is 
small for the purpose of our tests.

The amplitude difference between PN and NR waveforms 
is computed at the same connecting-point GW frequency. There is another effect which causes  
a jump in the hybrid-waveform amplitude. This is a small frequency 
difference between PN and NR waveforms at the connecting point. 
All our NR waveforms contain small eccentricities~\cite{BCP, Goddlong}. As a consequence, the
frequency evolution $\omega(t)$ oscillates.  To reduce this effect we
follow what is done in Ref.~\cite{Goddlong} and fit the frequency to a
monotonic quartic function.  When building the hybrid waveform, we
adjust the PN frequency to match the quartic fitted frequency (instead
of the oscillatory, numerical frequency) at the connecting
point. Since the restricted PN amplitude is proportional to
$\omega^{2/3}(t)$ [see Eq.~\eqref{quadstrain}], this slight difference
between $\omega$s at the connecting point creates another difference
between the NR and PN amplitudes.  Nevertheless, this difference is
usually smaller (for Goddard's waveform) or comparable (for Pretorius') to the
amplitude difference discussed above.
\begin{figure*}
\begin{center}
\includegraphics[width=0.95\textwidth]{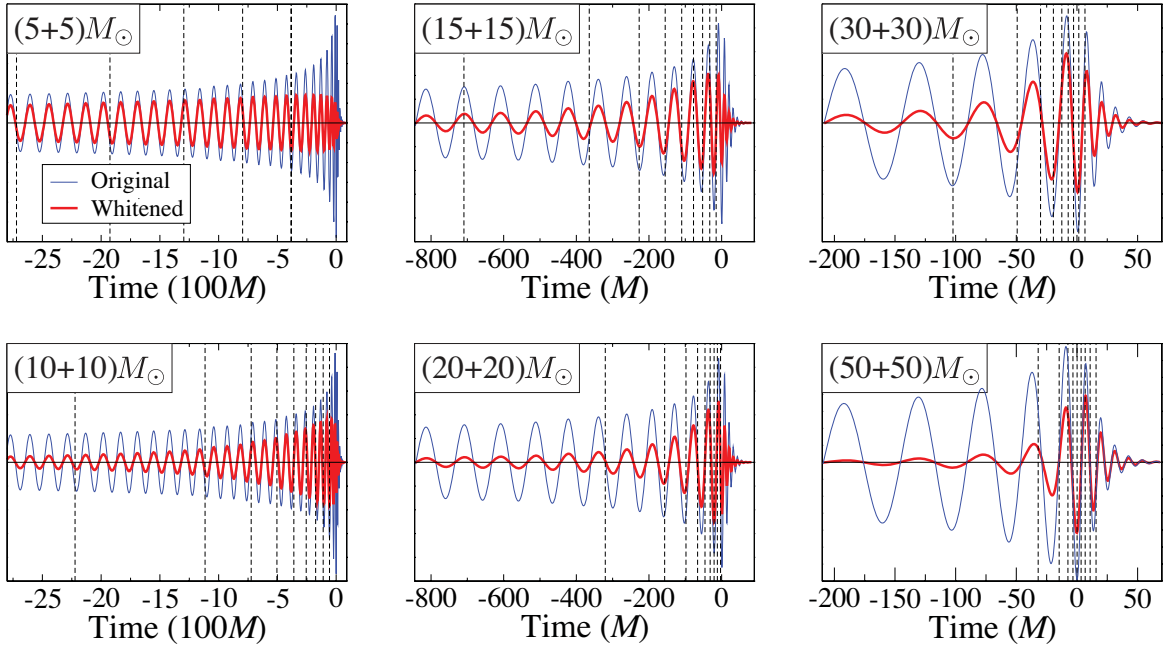}
\caption{Distribution of GW signal power. In each panel, we plot a hybrid waveform (a 
Tpn waveform stitched to the Goddard waveform) in both its original 
form (blue curve) and its ``whitened'' form (red curve) \cite{DIS00}. 
We show waveforms from six binary systems with total masses $10M_\odot$ $20M_\odot$, $30M_\odot$, 
$40M_\odot$, $60M_\odot$ and $100M_\odot$. The vertical lines divide the waveforms into segments, 
where each segment contributes $10\%$ of the total signal power.
\label{fig:SNRdist}}
\end{center}
\end{figure*}

\subsection{Distribution of signal power in gravitational waveforms}
\label{sec3.2}

To better understand the results of the FFs between hybrid waveforms, 
we want to compute how many {\it significant} GW cycles are in the 
LIGO frequency band. By significant GW cycles we 
mean the cycles that contribute most 
to the signal power, or to the SNR of the filtered signal. 
Since GW frequencies are scaled by the total binary mass,  
the answer to this question depends on both the PSD and the
total mass of a binary.

In Fig.~\ref{fig:SNRdist}, we show the effect of the LIGO PSD on the
distribution of signal power for several waves emitted by coalescing
binary systems with different total masses.  In each panel, we plot a
hybrid waveform (a Tpn waveform stitched to the Goddard waveform) in
both its original form and its ``whitened'' form~\cite{DIS00}. The whitened
waveform is generated by Fourier-transforming the original waveform
into the frequency domain, rescaling it by a factor of
$1/\sqrt{S_h(f)}$, and then inverse-Fourier-transforming it back to
the time domain. The reference time $t=0$  is the peak in the
amplitude of the unwhitened waveform.
The amplitude of a segment of the whitened waveform
indicates the relative contribution of that segment to the signal
power and takes into account LIGO's PSD.  Both waveforms are
plotted with arbitrary amplitudes, and the unwhitened one always has
the larger amplitude. The absolute amplitude of a waveform, or
equivalently the distance of the binary, is not relevant in these
figures unless the redshift $z$ becomes significant. In this case 
the mass of the binary is the redshifted mass $(1+z)M$. 
Vertical lines in each figure divide a waveform into segments,
where each segment contributes $10\%$ of the total signal power. In
each plot, except for the $10M_\odot$-binary one, we show all 9
vertical lines that divide the waveforms into 10 segments. In the
$10M_\odot$-binary plot we omit the early part of the inspiral phase that
accounts for $50\%$ of the signal power, as it would be too
long to show. 

The absolute time-scale of a waveform increases linearly with total mass $M$;
equivalently the waveform is shifted toward lower frequency
bands. For a $M=10M_\odot$ binary, the long inspiral stage generates
GWs with frequencies spanning the most sensitive part of the LIGO band, around
150Hz, while for an $M=100M_\odot$ binary, only the merger signal contributes 
in this band. Thus, for low-mass binary systems, most of the 
contribution to the signal power comes from the long
inspiral stage of the waveform, while for high-mass binary systems most
of the contribution comes from the late inspiral, merger, and ring-down
stages. Understanding quantitatively the distribution of signal power 
will let us deduce how many, and which, GW cycles are significant 
for the purpose of data analysis. We need accurate waveforms
from either PN models or NR simulations for at least those
significant cycles.

From Fig.~\ref{fig:SNRdist} we conclude that: 
\begin{itemize}
\item For an $M=10M_\odot$ binary, the last 25 inspiral cycles, plus
  the merger and ring-down stages of the waveform contribute only $50\%$
  of the signal power, and we need 80 cycles (not shown in the figure) of accurate inspiral
  waveforms to recover $90\%$ of the signal power. For an $M=20M_\odot$ 
  binary, the last 23 cycles, plus the merger and ring-down stages of the waveform contribute $>90\%$ of the
  signal power, and current NR simulations can produce waveforms of such length;
\item For an $M=30M_\odot$ binary, the last 11 inspiral cycles, plus the merger
  and ring-down stages of the waveform contribute $>90\%$ of the
  signal power, which means that, for binary systems with total masses
  higher than $30M_\odot$, current NR simulations,
  e.g., the sixteen cycles obtained in Ref.~\cite{Goddlong}, can provide 
  long enough waveforms for a
  matched-filter search of binary coalescence, as also found
  in Ref.~\cite{baumgarte_et_al}; 
\item For an $M=100M_\odot$ binary, $>90\%$ of the signal power comes
  from the last inspiral cycle, merger and ring-down stages of the
  waveform, with two cycles dominating the signal power. It is thus possible 
  to identify this waveform as a ``burst'' signal. 
\end{itemize}
Similar analyses can be also done for advanced LIGO and VIRGO. 

\subsection{Comparing hybrid waveforms}
\label{sec3.3}

We shall now compute ${\rm FF}_0$s between hybrid waveforms.  We fix
the total mass of the equal-mass binary in each comparison, i.e., we
{\it do not} optimize over mass parameters, but only on phase and time. 
We use the mismatch, defined as $1-{\rm FF}_0$, to measure the
difference between waveforms and we compute them for LIGO, advanced
LIGO, and VIRGO. Note that by using ${\rm FF}_0$, we test the
closeness among hybrid waveforms that are generated from binary systems
with the same physical parameters; in other words, we test whether
the waveforms are accurate enough for the purpose of parameter
estimation, rather than for the sole purpose of detecting GWs.
In the language of Ref.~\cite{DIS98} we are studying the 
{\it faithfulness} of the PN templates~\footnote{Following Ref.~\cite{DIS98}, 
{\it faithful} templates are templates that have
large overlaps, say $ \gaq \, 96.5\%$, with the 
expected signal maximizing {\it only}
 over the initial phase and time of arrival. By contrast when 
the maximization is done {\it also} on the binary masses, 
the templates are called {\it effectual}.}.

\begin{figure*}
\begin{center}
\includegraphics[width=0.9\textwidth]{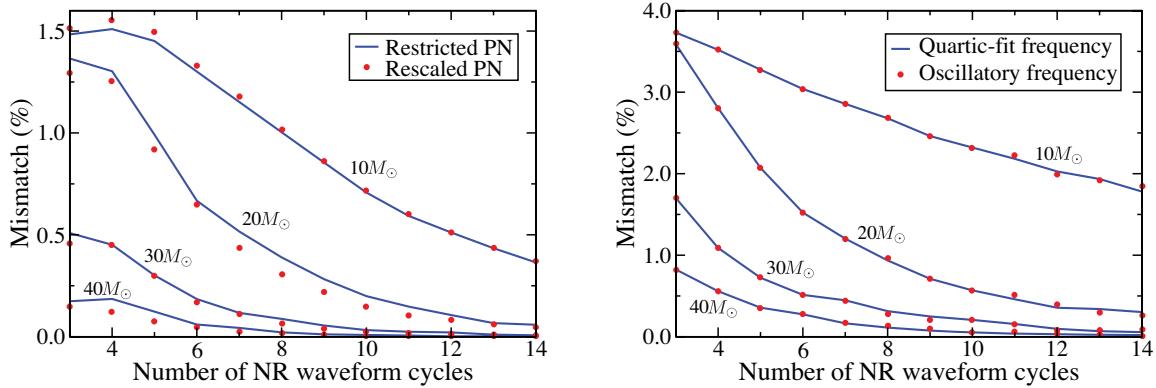}
\caption{We show the mismatch between hybrid waveforms as a function of the number 
of NR waveform cycles used to generate the hybrid waveforms. The LIGO PSD is used to evaluate the mismatches. In the 
left panel, we compare the Epn(3.5) and  Tpn(3.5) models. In the 
right panel, we compare the Tpn(3) and Tpn(3.5) models. 
From top to bottom, the four curves correspond 
to four equal-mass binary systems, with total 
masses $10M_\odot$, $20M_\odot$, $30M_\odot$, and $40M_\odot$. 
The dots show mismatches taken between hybrid 
waveforms that are generated with different methods. In the left panel, we adjust 
the amplitude of restricted PN waveforms, such that they connect smoothly 
in amplitude to NR waveforms. In the right panel, 
to set the frequency of PN waveforms at the joining point, 
we use the original orbital frequency, instead of the quartic fitted one. 
(See Sec.~\ref{sec3.1} for the discussion on amplitude scaling and frequency fitting). 
\label{fig:mmplot}}
\end{center}
\end{figure*}

\begin{figure*}
\begin{center}
\vspace{1cm}
\includegraphics[width=0.9\textwidth]{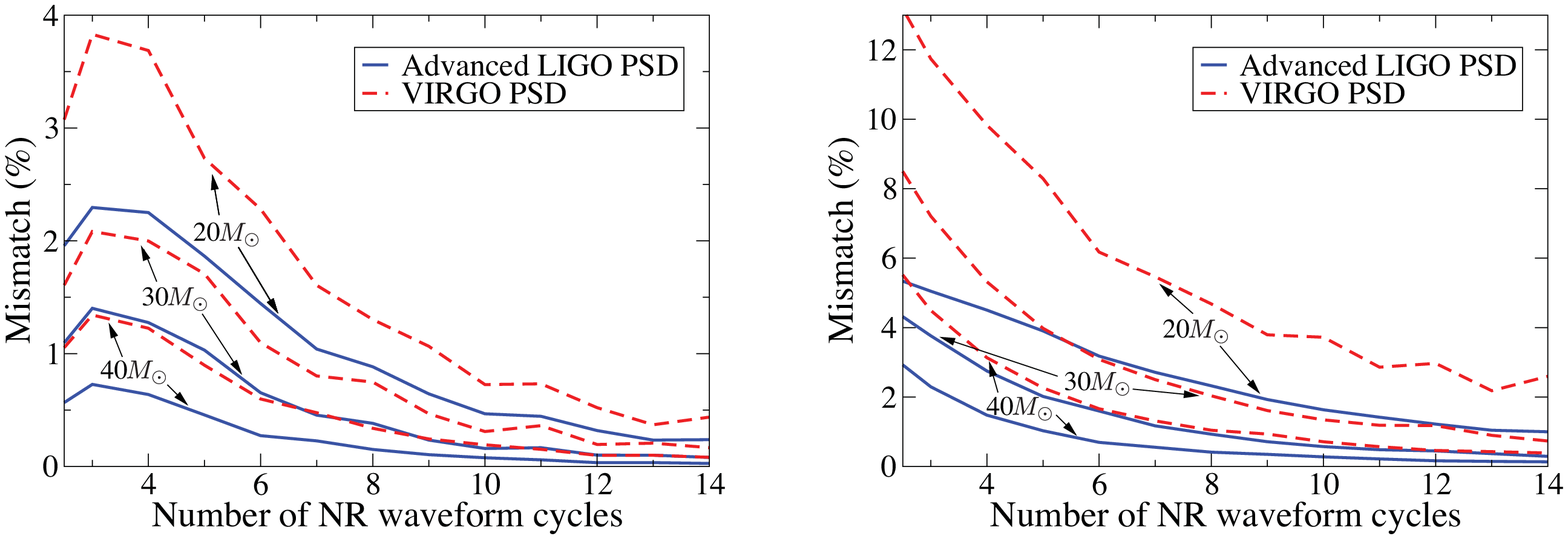}
\caption{Mismatch between hybrid waveforms as a function of the number 
of NR waveform cycles used to generate the hybrid waveforms. Following 
the settings of Fig.~\ref{fig:mmplot}, we show comparisons between Epn(3.5)
and  Tpn(3.5), and Tpn(3) and Tpn(3.5) models in the left 
and the right panels, respectively. The solid and dashed sets of curves are 
generated using the PSDs of advanced LIGO and VIRGO. In each set, 
 from top to bottom, the three 
curves correspond to three equal-mass binary systems, with total masses 
$20M_\odot$, $30M_\odot$, and $40M_\odot$. 
\label{fig:mmplotAV}}
\end{center}
\end{figure*}

  Since at late inspiral stages PN waveforms are partly replaced by
  NR waveforms, differences between hybrid waveforms from
  two PN models are smaller than those between pure PN waveforms. In
  general, the more NR cycles we use to generate hybrid waveforms, the
  less the difference is expected to be between these hybrid
  waveforms. This is evident in Figs.~\ref{fig:mmplot}, \ref{fig:mmplotAV} 
  where we show mismatches between hybrid
  waveforms for binary systems with different total masses as a function of
  the number of NR cycles $n$. Specifically, the mismatches are taken
  between two hybrid waveforms generated from the same NR waveform
  (from the Goddard group, taking the last $n$ cycles, plus
    merger and ring-down) and two different PN waveforms generated
  with the same masses. 

The mismatches are lower for binary systems with
higher total masses, since most of their signal power is concentrated in
the late cycles close to merger 
(see Fig.~\ref{fig:SNRdist}). Comparing results between LIGO,
advanced LIGO and VIRGO, we see that for the same waveforms the
mismatches are lowest when evaluated with the LIGO PSD, and highest
when evaluated with the VIRGO PSD. This is due to the much broader
bandwidth of VIRGO, especially at low frequency: the absolute
sensitivity is not relevant; only the shape of the PSD
matters. In VIRGO, the inspiral part of a hybrid waveform has higher
weighting in its contribution to the signal power. 
As already observed at the end of Sec.~\ref{sec2}, we can see
also that the difference between the Epn(3.5) and  Tpn(3.5) models is
smaller than that between the Tpn(3) and Tpn(3.5) models.


Figures~\ref{fig:mmplot}, \ref{fig:mmplotAV} show good agreement among
hybrid waveforms. In Sec.~\ref{sec4}, as a further confirmation 
of what was found in Refs.~\cite{BCP,Goddshort},
we shall see that PN waveforms from Tpn and Epn models have good 
agreement with the inspiral phase of the NR waveforms. 
Therefore, we argue that hybrid waveforms are likely 
to have high accuracy. In fact, for the late
evolution of a compact binary, where NR waveforms are available, the
PN waveforms are close to the NR waveforms, while for the early
evolution of the binary, where we expect the PN approximations to work
better, the PN waveforms (from Tpn and Epn models) are close to 
each other. Based on these observations, we draw the following conclusions 
for LIGO, advanced LIGO, and VIRGO data-analysis: 

\begin{itemize}
\item
For binary systems with total mass higher than $30M_\odot$, the current NR simulations of equal-mass
binary systems (16 cycles) are long enough to reduce mismatches between hybrid
waveforms generated from the three PN models to below
0.5\%. Since these FFs are achieved without optimizing the binary 
parameters, we conclude that for these
high-mass binary systems, the small difference between hybrid
waveforms indicates low systematic error in parameter estimation, 
i.e., hybrid waveforms are faithful \cite{DIS98}. 
\item 
For binary systems with total mass around $10 \mbox{--} 20M_\odot$, 16 cycles
  of NR waveforms can reduce the mismatch to below 3\%, which is
  usually set as the maximum tolerance for data-analysis purpose
  (corresponding to $\sim10\%$ loss in event rate). By a crude
  extrapolation of our results, we estimate that with $~30$ NR
  waveform cycles, the mismatch might be reduced to below $1\%$.
\item 
For binary systems with total mass lower than $10M_\odot$, the
  difference between the Tpn(3) and Tpn(3.5) models is substantial 
  for Advanced LIGO and VIRGO. Their mismatch can be $>4\%$ and $>6\%$ respectively
  (not shown in the figure). In this mass range, pursuing more NR
  waveform cycles in late inspiral phase does not help much, since the
  signal power is accumulated slowly over hundreds of GW cycles across
  the detector band. Nevertheless, here we give mismatches for 
  ${\rm FF}_0$s which are not optimized over binary masses. For the
  purpose of detection only, optimization over binary parameters leads 
  to low enough mismatches (see also the end of Sec.~\ref{sec2}). 
  In the language of Ref.~\cite{DIS98} hybrid waveforms for total 
  mass lower than $10M_\odot$  are {\it effectual} but not faithful.

\end{itemize}

\section{Matching numerical waveforms with post-Newtonian templates}
\label{sec4}

In this section, we compare the complete inspiral, merger and ring-down
waveforms of coalescing compact binary systems generated from NR simulations
with their best-match PN template waveforms. We also compare hybrid
waveforms with PN template waveforms for lower total masses,   
focusing on the late inspiral phase provided by the NR waveforms. We
test seven families of PN templates that either have been used in
searches for GWs in LIGO (see e.g., Refs.~\cite{LSC1, LSC2}), or are
promising candidates for ongoing and future searches with ground-based
detectors.  We evaluate the performance of PN templates by computing
the FFs maximized on phase, time and binary parameters. As we shall
see, for the hybrid waveforms of binary systems with total mass
$M\le30M_\odot$, both the time-domain families Tpn(3.5) and
Epn(3.5), which includes a superposition of three ring-down modes,  
perform well, confirming what found in 
Refs.~\cite{BCP,Goddshort}. The standard stationary-phase-approximated
(SPA) template family in the frequency domain has high FFs only for
binary systems with $M<20M_\odot$. After investigating in detail the GW phase 
in frequency domain, and having understood why it happens (see Sec.~\ref{sec4.2.2}), 
we introduce two modified SPA template families 
(defined in Sec.~\ref{sec4.2.2}) for binary systems with total mass
$M\ge30M_\odot$. Overall, for masses $M\ge30M_\odot$, 
the Epn(3.5) template family in the time domain and the two 
modified SPA template families in the frequency domain exhibit
the best-match performances.

\subsection{Numerical  waveforms and post-Newtonian templates}
\label{sec4.1}

For binary systems with total mass $M\ge30M_\odot$, the last 8--16 cycles 
contribute more than 80--90\% of the signal power, thus in this 
case we use only the NR waveforms. By contrast, for
binary systems with total mass $10\le M\le30M_\odot$, for which the merger
and ring-down phases of the waveforms contribute 
{\it only} $\sim 1\mbox{--}10\%$, we use the hybrid waveforms, 
generated by stitching Tpn waveforms to the Goddard NR waveforms. 

We want to emphasize that FFs computed for different target 
numerical waveforms can not directly be compared with each other. For
instance, the Goddard waveform is longer than the Pretorius waveform,
and the FFs are sometime slightly lower using the Goddard waveform. This is a completely artificial 
effect, due to the fact that it is much easier to tune 
the template parameters and obtain a large FF with a shorter 
target waveform than a longer one. 

\begin{table*}
\begin{tabular}{r||r|r|r}
& $(5+5)M_\odot$ & $(10+10)M_\odot$ & $(15+15)M_\odot$ \\
\hline\hline
Signal Power (\%) & (30, 0.2) & (80, 2) & (85, 10) \\
\hline
$\langle h^{\rm NR-hybr}, h^{\rm  Tpn(3.5)} \rangle $  & 0.9875 & 0.9527 & 0.8975 \\
($M/M_\odot, \eta$) & (10.18, 0.2422) & (19.97, 0.2500) & (29.60, 0.2499) \\
$M\omega_{\rm orb}$ & 0.1262 & 0.1287 & 0.1287 \\
\hline
$\langle h^{\rm NR-hybr}, h^{\rm  Epn(3.5)} \rangle $ & 0.9836 & 0.9522 & 0.9618 \\
($M/M_\odot, \eta$) & (10.15, 0.2435) & (19.90, 0.2500) & (29.49,0.2488) \\
($\epsilon_t, \epsilon_M, \epsilon_J$)(\%) & (-0.02, 12.19, 30.87) & (-0.02, 75.03, 95.00) & (0.05, 2.38, 92.06) \\
$M\omega_{\rm orb}$ & 0.1346 & 0.1345 & 0.1345 \\
\hline
$\langle h^{\rm NR-hybr}, h^{\rm  SPA_c(3.5)} \rangle $  & 0.9690 & 0.9290 & 0.8355 \\
($M/M_\odot, \eta$) & (10.16, 0.2432) & (19.93, 0.2498) & (29.08, 0.2500) \\
($f_{\rm cut}/{\rm Hz}$) & $1566.8$ & 263.9 & 529.6 \\
\end{tabular}

\vspace{0.5cm}

\caption{FFs between hybrid waveforms [Tpn(3.5) waveform stitched to the Goddard waveform] and PN templates. 
In the first row, the two numbers in parentheses are the percentages of the signal-power contribution from the 16 
inspiraling NR cycles and the NR merger/ring-down cycles. (The separation between inspiral and merger/ring-down
is obtained using the EOB approach as a guide, i.e., we match the Epn(3.5) model and use the 
EOB light-ring position as the beginning of the merger phase.) 
In the PN-template rows, the first number in each block is the FF, and the numbers in parentheses are template 
parameters that achieve this FF. The last number in each block of the Tpn(3.5) and Epn(3.5) 
models is the ending orbital frequency of the best-match template. For the Epn model, the ending frequency 
is computed at the point of matching with the ring-down phase, around the EOB light ring.
\label{Tab1}}
\end{table*}
\begin{table*}
\begin{tabular}{r||r|r|r|r}
& $(15+15)M_\odot$ &$(20+20)M_\odot$ & $(30+30)M_\odot$ & $(50+50)M_\odot$ \\
\hline\hline
$\langle h^{\rm NR-Pretorius}, h^{\rm  Epn(3.5)} \rangle $  & 0.9616 & 0.9599 & 0.9602 & 0.9787 \\
($M/M_\odot, \eta$) & (27.93, 0.2384) & (35.77, 0.2426) & (52.27, 0.2370) & (96.60, 0.2386) \\
($\epsilon_t, \epsilon_M, \epsilon_J$)(\%) & (-0.08, 0.63, 99.70) & (-0.03, 0.48, 94.38) & (-0.12, 0.00, 64.14) & (0.04, 0.01, 73.01) \\
\hline
$\langle h^{\rm NR-Pretorius}, h^{\rm  SPA_c^{\rm ext}(3.5)} \rangle $ 
& 0.9712 & 0.9802 & 0.9821 & 0.9722 \\
($M/M_\odot, \eta$) & (19.14, 08037) & (24.92, 0.9097) & (36.75, 0.9933) & (58.06, 0.9986) \\
($f_{\rm cut}/{\rm Hz}$) & (589.6) & (476.9) & (318.9) & (195.9) \\
\hline
$\langle h^{\rm NR-Pretorius}, h^{\rm  SPA_c^{\mathcal{Y}}(4)} \rangle $ & 0.9736 & 0.9824 & 0.9874 & 0.9851 \\
($M/M_\odot, \eta$) & (29.08, 0.2460) & (38.63, 0.2461) & (57.58, 0.2441) & (96.55, 0.2457) \\
($f_{\rm cut}/{\rm Hz}$) & (666.5) & (501.2) & (332.5) & (199.4) \\
\hline
$\langle h^{\rm NR-Pretorius}, h^{\rm BCV} \rangle $ & 0.9726 & 0.9807 & 0.9788 & 0.9662 \\
($\psi_0/10^4, \psi_1/10^2$) & (2.101, 1.655) & (1.178, 1.744) & (0.342, 2.385) & (-0.092, 3.129) \\ 
($10^2\alpha, f_{\rm cut}/{\rm Hz}$) & (-1.081, 605.5) & (-0.834, 461.7) & (0.162, 320.4) & (1.438, 204.3) \\
\hline
$\langle h^{\rm NR-Pretorius}, h^{\rm BCV_{\rm impr}} \rangle $ & 0.9727 & 0.9807 & 0.9820 & 0.9803 \\
($\psi_0/10^4, \psi_1/10^2$) & (2.377, 0.930) & (1.167, 1.762) & (0.431, 2.077) & (-0.109, 3.158) \\ 
($10^2\alpha, f_{\rm cut}/{\rm Hz}$) & (-3.398, 571.9) & (-2.648, 458.3) & (-1.196, 319.1) & (-3.233, 196.0) \\
\end {tabular}

\vspace{0.5cm}

\begin{tabular}{r||r|r|r|r}
& $(15+15)M_\odot$ &$(20+20)M_\odot$ & $(30+30)M_\odot$ & $(50+50)M_\odot$ \\
\hline\hline
$\langle h^{\rm NR-Goddard}, h^{\rm  Epn(3.5)} \rangle $ & 0.9805 & 0.9720 & 0.9692 & 0.9671 \\
($M/M_\odot, \eta$) & (29.25, 0.2435) & (38.27, 0.2422) & (56.66, 0.2381) & (83.52, 0.2233) \\
($\epsilon_t, \epsilon_M, \epsilon_J$)(\%) & (0.05, 0.03, 99.90) & (0.05, 0.27, 99.17) 
& (0.09, 0.01, 54.56) & (0.10, 1.71, 79.75) \\
\hline
$\langle h^{\rm NR-Goddard}, h^{\rm  SPA_c^{\rm ext}(3.5)} \rangle $ & 0.9794 & 0.9785 & 0.9778 & 0.9693 \\
($M/M_\odot, \eta$) & (21.41, 0.5708) & (27.27, 0.6695) & (37.67, 0.9911) & (60.90, 0.9947) \\
($f_{\rm cut}$/{\rm Hz}) & (552.7) & (444.4) & (318.5) & (191.7) \\
\hline
$\langle h^{\rm NR-Goddard}, h^{\rm  SPA_c^{\mathcal{Y}}(4)} \rangle $ & 0.9898 & 0.9905 & 0.9885 & 0.9835 \\
($M/M_\odot, \eta$) & (30.28, 0.2456) & (40.23, 0.2477) & (60.54, 0.2455) & (100.00, 0.2462) \\
($f_{\rm cut}$/{\rm Hz}) & (674.6) & (506.6) & (330.5) & (195.0) \\
\hline
$\langle h^{\rm NR-Goddard}, h^{\rm BCV} \rangle $ & 0.9707 & 0.9710 & 0.9722 & 0.9692 \\
($\psi_0/10^4, \psi_1/10^2$) & (3.056, -1.385) & (1.650, -0.091) & (0.561, 1.404) & (-0.113, 3.113) \\ 
($10^2\alpha, f_{\rm cut}$/{\rm Hz}) & (0.805, 458.3) & (0.559, 412.6) & (0.218, 309.2) & (1.063, 198.7) \\
\hline
$\langle h^{\rm NR-Goddard}, h^{\rm BCV_{\rm impr}} \rangle $ & 0.9763 & 0.9768 & 0.9782 & 0.9803 \\
($\psi_0/10^4, \psi_1/10^2$) & (2.867, -0.600) & (1.514, 0.448) & (0.555, 1.425) & (-0.165, 3.373) \\ 
($10^2\alpha, f_{\rm cut}$/{\rm Hz}) & (0.193, 578.0) & (-1.797, 441.1) & (-4.472, 308.1) & (-4.467, 193.4) \\
\end {tabular}

\vspace{0.5cm}

\caption{FFs between NR waveforms and PN templates which include merger and ring-down phases. 
The upper table uses Pretorius' waveform, and the lower table uses 
Goddard's high-resolution long waveform. 
The first number in each block is the FF, and numbers in parentheses are template parameters 
that achieve this FF.
\label{Tab2}}
\end{table*}

We consider seven PN template families. The two time-domain families introduced in Sec.~\ref{sec2} are:

\vspace{0.5cm}

$\bullet$  Tpn(3.5)~\cite{bcv1,bcv2}: The inspiral Taylor model.

\vspace{0.5cm}

$\bullet$ Epn(3.5)~\cite{BD1,BD2,DJS,DIS98,BCP}: The EOB model which includes 
a superposition of three quasi-normal modes (QNMs) of the final BH. These
are labeled by three integers $(l,m,n)$~\cite{QNM}: the least damped 
QNM $(2,2,0)$ and two overtones $(2,2,1)$ and $(2,2,2)$. The ring-down waveform is given as: 
\beq
h_{\rm QNM}(t) = \sum_{n=0}^{2} A_ne^{-(t-t_{\rm end})/\tau_{\rm 22n}} \,
\cos\left[\omega_{\rm 22n} (t-t_{\rm end})+\phi_n\right]\,,
\label{qnmw}
\eeq
where $\omega_{lmn}$ and $\tau_{lmn}$ are the frequency and decay time
of the QNM $(l,m,n)$, determined by the mass $M_f$ and spin $a_f$
 of the final BH.
The quantities $A_n$ and $\phi_n$ in Eq.~(\ref{qnmw}) 
are the amplitude and phase of the QNM $(2,2,n)$. They 
are obtained by imposing the continuity of $h_+$ and $h_\times$, and their
first and second time derivatives, at the time of matching $t_{\rm  match}$.  
Besides the mass parameters, our Epn model contains three
other {\it physical} parameters: $\epsilon_t$, $\epsilon_M$ and
$\epsilon_J$. The parameter $\epsilon_t$ takes into account possible
differences between the time $t_{\rm end}$ at which the EOB models
end and the time $t_{\rm match}$ at which the matching to ring-down
is done. More explicitly, we set $t_{\rm match}=(1+\epsilon_t)t_{\rm
  end}$, and if $\epsilon_t>0$, we extrapolate the EOB evolution,
and set an upper limit for the $\epsilon_t$ search where the
extrapolation fails. The parameters $\epsilon_M$ and $\epsilon_J$
describe possible differences between the values of the mass $M_{\rm
  end} \equiv E_{\rm end}$ and angular momentum $\hat{a}_{\rm end} \equiv
J_{\rm end}/M_{\rm end}^2$ at the end of the EOB inspiral and the
final BH mass and angular momentum. (The end of the EOB inspiral 
occurs around the EOB light-ring.) The differences are due to the
fact that the system has yet to release energy and angular momentum
during the merger and ring-down phase before settling down to the
stationary BH solution. If the total binary mass and angular momentum
at the end of the EOB inspiral are $M_{\rm end}$ and $J_{\rm end}$, we
set the total mass and angular momentum of the final stationary BH to
be $M_f=(1-\epsilon_M) M_{\rm end}$ and $J_f=(1-\epsilon_J) J_{\rm
  end}$, and use $a_f\equiv J_f/M_f$ to compute $\omega_{lmn}$ and
$\tau_{lmn}$. We consider the current Epn model with three
parameters $\epsilon_t$, $\epsilon_M$ and $\epsilon_J$, 
as a first attempt to build a physical EOB model for matching 
{\it coherently} the inspiral, merger and ring-down phases. 
Since the $\epsilon$-parameters are related to physical
quantities, e.g., the loss of energy during ring-down, 
they are functions of the initial physical parameters of the binary, such 
as masses, spins, etc. In the near future we expect to be able to 
fix the $\epsilon$-values by comparing NR and (improved) EOB waveforms 
for a large range of binary parameters.

We also consider five frequency-domain models, in which two (modified SPA models) 
are introduced later in Sec.~\ref{sec4.2.2}, and three are introduced here:
\vspace{0.5cm}

$\bullet$  SPA$_c$(3.5)~\cite{SPA}: SPA$_c$ PN model with an appropriate cutoff
  frequency $f_{\rm cut}$~\cite{bcv1,bcv2};

\vspace{0.5cm}

$\bullet$ BCV~\cite{bcv1}: BCV model with an amplitude correction term
  ($1-\alpha f^{2/3}$) and an appropriate cutoff frequency $f_{\rm cut}$.

\vspace{0.5cm}

$\bullet$ BCV$_{\rm impr}$~\cite{bcv1}: Improved BCV model with an
  amplitude correction term ($1-\alpha f^{1/2}$) and an appropriate
  cutoff frequency $f_{\rm cut}$. We include this improved BCV model because
  Ref.~\cite{BCP} found a deviation of the Fourier-transform amplitude
  from the Newtonian prediction $f^{-7/6}$ during the merger and
  ring-down phases (see Fig. 22 of Ref.~\cite{BCP}). Here we shall
  assume $n=-2/3$ in the $f^n$ power law to get the ($1-\alpha
  f^{1/2}$) form of the amplitude correction.  While it was
  found~\cite{BCP} that the value of $n$ is close to $-2/3$ for the
  $l=2,m=2$ waveform, this value varies slightly if other multiple moments are
  included and if binary systems with different mass ratios are considered.
  Finally, the $\alpha$ parameter is expected to be negative, but in
  our actual search it can take both positive and negative values.

\subsection{Discussion of fitting-factor results}
\label{sec4.2}
 
In Table~\ref{Tab1}, we list the FFs for hybrid target waveforms and three 
PN template families: Tpn(3.5), Epn(3.5), and SPA$_c$(3.5), together with the template parameters 
at which the best match is obtained. As shown in the first row, in this relatively low-mass range, 
i.e. $10M_\odot<M<30M_\odot$, the merger/ring-down phases of the waveforms contribute only a small 
fraction of the total signal power, while the last 16 inspiraling cycles of the NR waveform 
contribute a significant fraction. 
Therefore, confirming recent claims by Refs.~\cite{BCP,Goddshort}, 
we can conclude that the PN template families Tpn(3.5) and Epn(3.5) 
have good agreement with the inspiraling NR waveforms. 
The Tpn(3.5) model gives a low FF for $M=30M_\odot$ because 
for these higher masses the merger/ring-down phases, which the Tpn model does not
include, start contributing to the signal power.
Note that both time-domain templates give fairly good estimates of the mass parameters. 
The SPA$_c$(3.5) template family gives FFs that drop substantially when the total 
binary mass increases from $10M_\odot$ to $30M_\odot$, indicating that this 
template family can only match the early, less relativistic inspiral phase of the hybrid waveforms. 
Nevertheless, it turns out that by slightly modifying the SPA waveform we   
can match the NR waveforms with high FFs (see Sec.~\ref{sec4.2.2}). 

\begin{figure}
\includegraphics[width=0.5\textwidth]{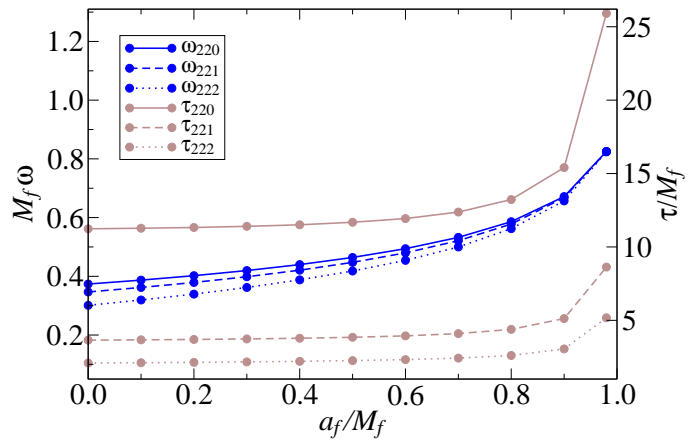}
\caption{Frequencies and decay times of the least damped QNM 220, and two overtones 221 and 222. 
The scales of the frequency and the decay time are listed on the left and right sides of the plot, respectively.
\label{fig:qnmfreq}}
\end{figure}

In Table~\ref{Tab2}, we list the FFs for full NR waveforms and five PN template families: Epn(3.5), 
SPA$_c^{\rm ext}$(3.5), SPA$_c^{\mathcal{Y}}$(4), BCV, and BCV$_{\rm impr}$, together with the template parameters 
at which the best match is obtained. The SPA$_c^{\rm ext}$(3.5) and SPA$_c^{\mathcal{Y}}$(4) families are
modified versions of the SPA family, defined in Sec.~\ref{sec4.2.2}.

We shall investigate these results in more detail in the following sections.

\subsubsection{Effective-one-body template performances}

The Epn model is the only available time-domain model that explicitly includes ring-down waveforms. 
It achieves high ${\rm FFs}\ge0.96$  for all target waveforms, confirming 
the necessity of including ring-down modes and proving that the inclusion of three QNMs with three 
tuning parameters $\epsilon_t$, $\epsilon_M$ and $\epsilon_J$ is sufficient for detection. 
As we see in Table~\ref{Tab2}, the values of the tuning parameters $\epsilon_M$ 
and $\epsilon_J$, where the FFs are achieved, are 
different from their physical values. For reference, the Goddard numerical simulation predicts 
$M_f\simeq 0.95 M$ and $\hat{a}_f\equiv J_f/M^2_f\simeq 0.7$~\cite{Goddlong}, 
and  Epn(3.5) predicts $M_{\rm end}= 0.967$ and $\hat{a}_{\rm end}\equiv J_{\rm end}/M^2_{\rm end}=0.796$, 
so the two tuning parameters should be $\epsilon_M \simeq 1.75\%$ and $\epsilon_J \simeq 11\%$. 
In our search, e.g., for $M=30M_\odot$, $\epsilon_J$ tends to be tuned to its lowest possible value and 
$\epsilon_t$ tends to take its highest possible value, indicating that pushing the end of 
the  Epn(3.5) inspiral to a later time gives higher FFs. 

\begin{figure*}
\begin{center}
\includegraphics[width=1.0\textwidth]{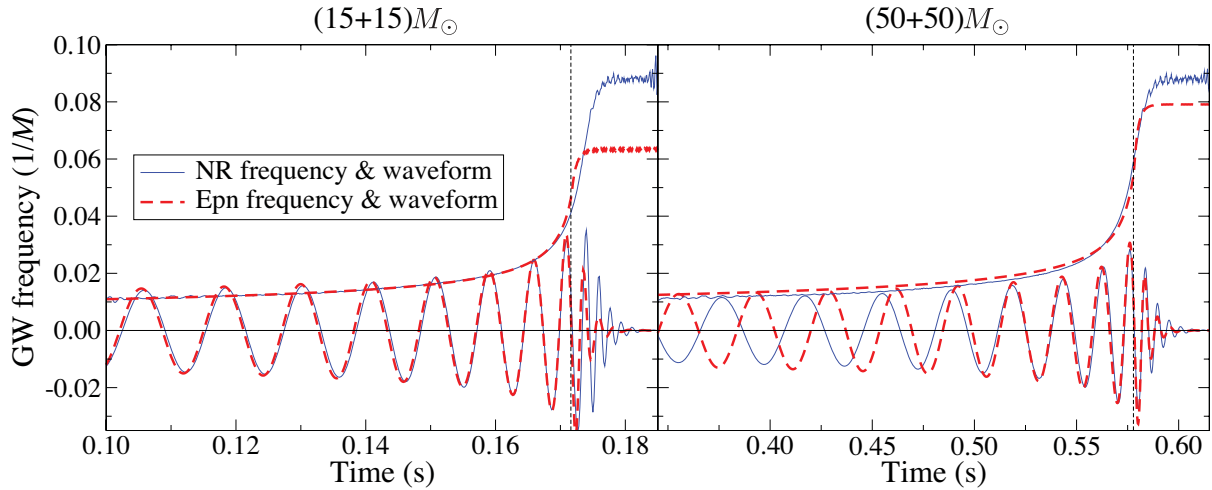}
\caption{Frequency evolution of waveforms from the Epn(3.5) model, and the NR simulations of the Goddard group. In the left and right panels, we show frequency evolutions for two equal-mass binary systems with total mass $30M_\odot$ and $100M_\odot$. In each panel, there are two nearly monotonic curves and two oscillatory curves, where the former are frequency evolutions and the latter are binary coalescence waveforms. The solid curves (blue) are from the NR simulations, while the dashed curves (red) are from the Epn(3.5) model. The vertical line in each plot shows the position where the three-QNM ring-down waveform is attached to the EOB waveform. 
  \label{fig:freqevol}}
\end{center}
\end{figure*}

Since the parameters $\epsilon_M$ and $\epsilon_J$ depend on the QNM frequency and 
decay time, we show in Fig.~\ref{fig:qnmfreq} how $\omega_{lmn}$ and $\tau_{lmn}$ vary 
as functions of $a_f$~\cite{QNM} for the three modes used in the Epn(3.5) model. 
The frequencies $\omega_{lmn}$ of the three modes are not really different, and grow monotonically with 
increasing $a_f$. The decay times $\tau_{lmn}$, although different for the three modes, also 
grow monotonically with increasing $a_f$. Thus, the huge loss of angular momentum $\epsilon_J$, 
or equivalently the small final BH spin required in the Epn(3.5) model to achieve high FFs, indicates 
that {\it low} ring-down frequencies and/or {\it short} decay times are needed for this model 
to match the numerical merger and ring-down waveforms.

In Fig.~\ref{fig:freqevol}, we show Goddard NR and Epn(3.5) waveforms, as well as their 
frequency evolutions, for two equal-mass binary systems with total masses $30M_\odot$ and $100M_\odot$. 
In the low-mass case, i.e., $M=30M_\odot$, since the inspiral part contributes most of the SNR, 
the Epn(3.5) model fits the frequency and phase evolution of the NR inspiral well, with the 
drawback that at the joining point the EOB frequency is substantially higher than that of 
the NR waveform. Then, in order to fit the early ring-down waveform which has higher amplitude, 
the tuning parameters have to take values in Table~\ref{Tab2} such that the ring-down 
frequency is small enough to get close to the NR frequency during {\it early} ring-down stage, as 
indicated in Fig.~\ref{fig:freqevol}. The late ring-down waveform does not contribute much to the SNR, and thus it is not too surprising
that waveform optimizing the FF does not adequately represent this part of the NR waveform.
In the higher mass case, $M=100M_\odot$, the Epn(3.5) model gives a much better, though not
perfect, match to the merger and ring-down phases of the NR waveform, at the expense of
misrepresenting the early inspiral part. Again, this is not unexpected considering
that in this mass range the merger and ring-down waveforms dominate the contribution to the SNR.

Comparing the two cases discussed above, we can see that with the current procedure of matching the
inspiral and ring-down waveforms in the EOB approach it is not possible to obtain a perfect 
match with the entire NR waveform. However, due to the limited detector sensitivity bandwidth, 
the FFs are high enough for detection. The large systematic error in estimating the 
physical parameters will be overcome by improving the EOB matching procedure 
during the inspiral part, and also by fixing the $\epsilon$-parameters to 
physical values obtained by comparison with numerical simulations. 

Finally, in Figs.~\ref{fig:SPAfcfreq}, \ref{fig:SPAfcfreq100} we 
show the frequency-domain amplitude and phase of the NR and EOB waveforms. 
Quite interestingly, we notice that the inclusion of three ring-down modes 
reproduce rather well the {\it bump} in the NR frequency-domain amplitude. 
The EOB frequency-domain phase also matches the NR one very well.

\subsubsection{Stationary-phase-approximated template performances}
\label{sec4.2.2}

\begin{figure}
\includegraphics[width=0.5\textwidth]{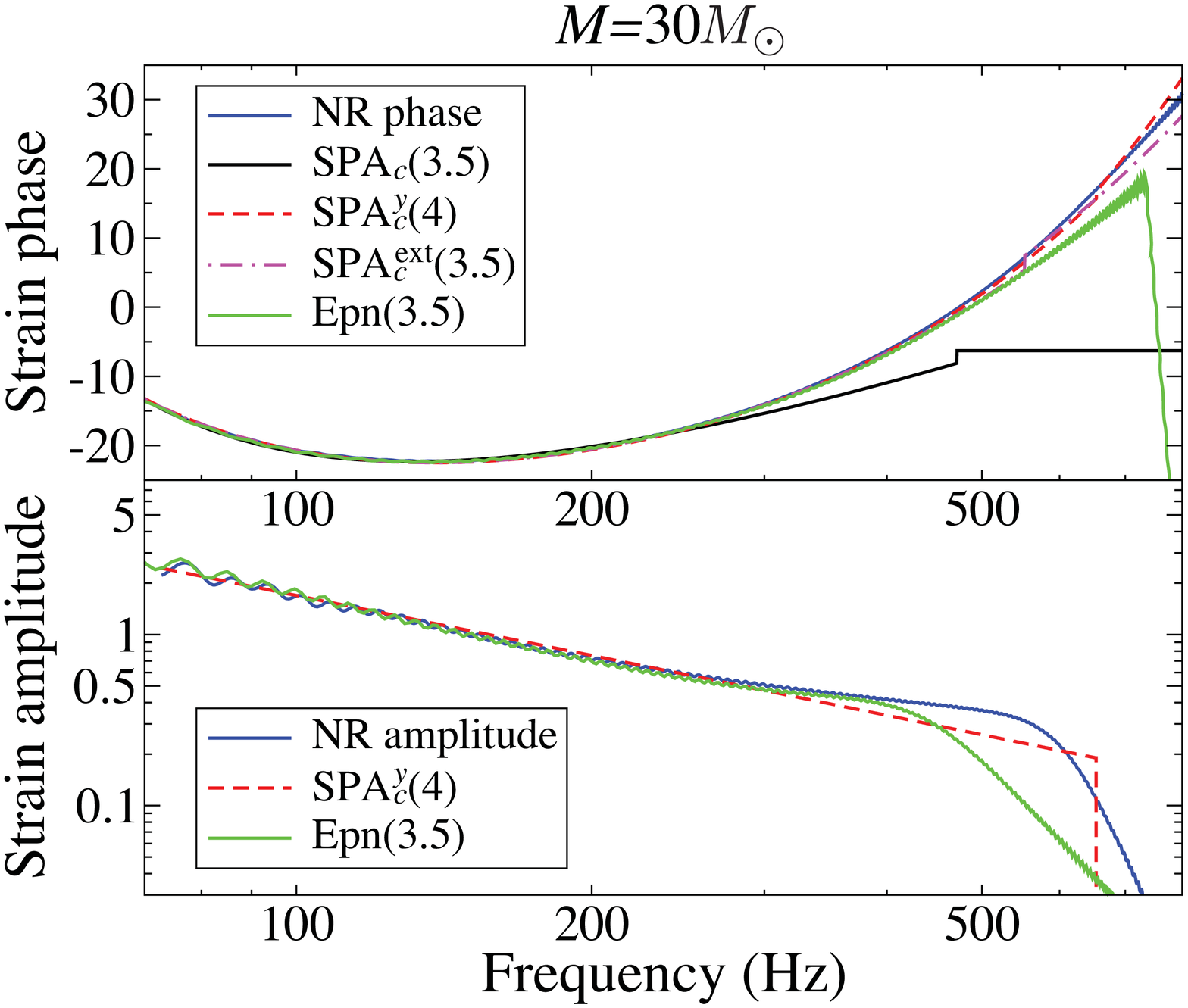}
\caption{For $M=30M_\odot$ equal-mass binary systems, we compare the phase and amplitude of the frequency-domain waveforms from the SPA$_c$ models and NR simulation (Goddard group). 
We also show the amplitude of the waveform from the Epn(3.5) model.
\label{fig:SPAfcfreq}}
\end{figure}

\begin{figure}
\includegraphics[width=0.5\textwidth]{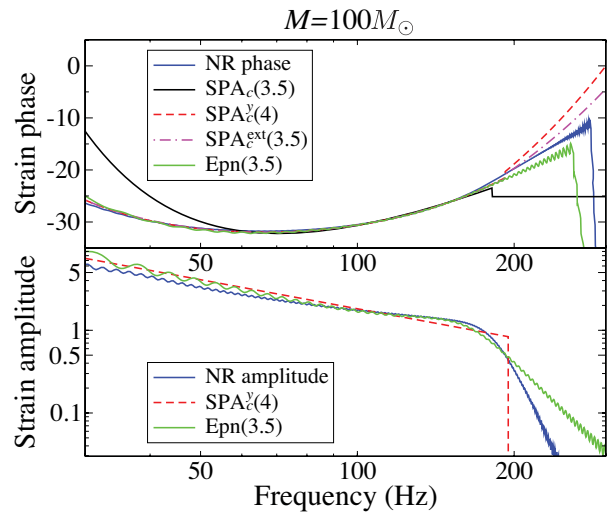}
\caption{For $M=100M_\odot$ equal-mass binary systems, we compare the phase and amplitude of the frequency-domain waveforms from the SPA$_c$ models and NR simulation (Goddard group). 
We also show the amplitude of the waveform from the Epn(3.5) model.
\label{fig:SPAfcfreq100}}
\end{figure}

Figures~\ref{fig:SPAfcfreq}, \ref{fig:SPAfcfreq100} also show the frequency-domain phases and amplitudes 
for the best-match SPA$_c(3.5)$ waveforms. We see that at high frequency the NR and 
SPA$_c$(3.5) phases rise with different slopes
~\footnote{By looking in detail at the PN terms in the SPA$_c$(3.5) phase, 
we find that the difference in slope is largely due to the logarithmic term at 2.5PN order.}. 
Based on this observation we introduce two modified SPA models:
\vspace{0.5cm}

$\bullet$ SPA$_c^{\rm ext}$(3.5): SPA$_c$ PN model with unphysical values of $\eta$ 
and an appropriate cutoff frequency $f_{\rm cut}$. 
The range of the symmetric mass-ratio $\eta=m_1m_2/(m_1+m_2)^2$ 
is extended from its physical range $0\sim0.25$ to the unphysical range $0\sim1$.

\vspace{0.5cm}

$\bullet$ SPA$_c^{\mathcal{Y}}$(4): SPA$_c$ PN model with an {\it ad hoc}  4PN order term 
in the phase, and an appropriate cutoff frequency $f_{\rm cut}$. 
The phase of the SPA model is known up to the 3.5PN order (see, e.g., Eq. (3.3) of Ref.~\cite{SPA}):
\beq
\psi(f)=2\pi ft_0-\phi_0-\frac{\pi}{4}+\frac{3}{128\eta v^5}\sum_{k=0}^N\alpha_kv^k\,,
\eeq
where $v=(\pi Mf)^{1/3}$. The PN coefficients $\alpha_k$s, $k=0,\dots,N$, 
(with $N=7$ at 3.5PN order) are given by Eqs. (3.4a), (3.4h) of Ref.~\cite{SPA}. 
We add the following term at 4PN order:
\beq
\alpha_8=\mathcal{Y}\log v\,,
\eeq
where $\mathcal{Y}$ is a parameter which we fix by imposing high matching performances  
with NR waveforms. Note that a constant term in $\alpha_8$ only adds a 4PN order term that is linear in $f$, 
which can be absorbed into the $2\pi ft_0$ term. Thus, to obtain a nontrivial effect, we 
need to introduce a logarithmic term. The coefficient  $\mathcal{Y}$ could in principle depend on 
$\eta$. We determine $\mathcal{Y}$ by optimizing the FFs of equal and unequal 
masses. We find that in the equal-mass case $\mathcal{Y}$ does not depend significantly on the binary total mass 
and is given by $\mathcal{Y}=3923$. The latter is also close to the best match value obtained 
for unequal masses. More specifically, it is within 4.5\% for binary systems of mass ratio 
$m_2/m_1=2$. To further explore the dependence of $\mathcal{Y}$ on $\eta$, we need a larger sample 
of waveforms for unequal-mass binary systems~\footnote{Note that the {\it auxiliary phase} 
introduced in Eq. (239) of Ref.~\cite{LR} also gives rise to a term in the SPA phase 
of the kind $f\,\log v$, except an order of magnitude smaller than $\mathcal{Y}$.}.
As seen in Table~\ref{Tab2}, the two modified SPA$_c$ 
template families have FF$>0.97$ (except for one 0.9693) for all target waveforms, 
even though no explicit merger or ring-down phases are included in the waveform. 
The SPA$_c^{\mathcal{Y}}$(4) model provides also a really good estimation of parameters.

\begin{figure*}
\begin{center}
\includegraphics[width=1.0\textwidth]{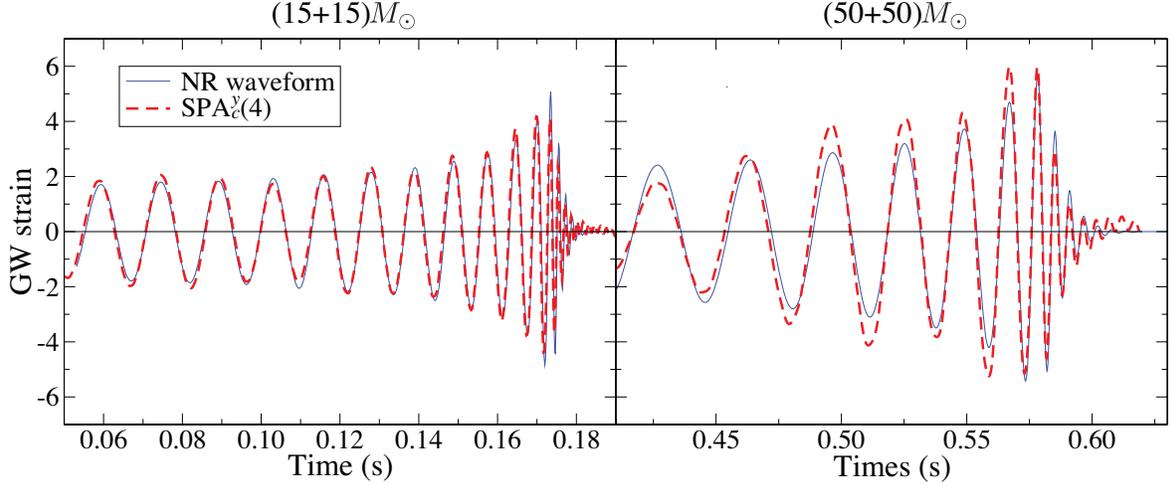}
\caption{Binary coalescence waveforms from the SPA$_c^{\mathcal{Y}}$(4) model, and the NR simulations 
of the Goddard group. In the left and right panels we show waveforms for two equal-mass 
binary systems with total mass $30M_\odot$ and $100M_\odot$. The solid lines show the waveforms from 
the NR simulation, and the dashed lines give the best-matching waveforms from the SPA$_c^{\mathcal{Y}}$(4) model.
\label{fig:SPAfc}}
\end{center}
\end{figure*}

In Fig.~\ref{fig:SPAfc} we plot Goddard NR and SPA$_c^{\mathcal{Y}}$(4) waveforms for two equal-mass binary systems with total 
masses $M=30M_\odot$ and $M=100M_\odot$. We can clearly see tail-like ring-down waveforms at the end of 
the SPA$_c^{\mathcal{Y}}$(4) waveforms, which result from the inverse Fourier transform of frequency domain waveforms 
that have been cut at $f=f_{\rm cut}$. This well-known feature is called the Gibbs phenomenon. 
At first glance, it may appear surprising that the often inconvenient Gibbs phenomenon~\cite{DIS00} can 
provide reasonable ring-down waveforms in the time domain. However, by looking at 
the spectra of these waveforms in the frequency domain (see the amplitudes 
in Figs.~\ref{fig:SPAfcfreq} \& \ref{fig:SPAfcfreq100}), we see that the SPA$_c^{\mathcal{Y}}$(4) cuts off 
at the frequency $f_{\rm cut}$ (obtained from the optimized FF) where the NR 
spectra also start to drop. Thus, even though the frequency-domain SPA$_c$ 
waveforms are discontinuous, while the frequency-domain NR waveforms are continuous (being combinations of Lorentzians), 
the SPA$_c$ time-domain waveforms contain tails with frequencies and decay rates similar to the NR 
ring-down modes. We expect that the values of the cutoff frequency $f_{\rm cut}$ at which the FFs are 
maximized are well-determined by the highest frequency of the NR waveforms, 
i.e. by the frequency of the fundamental QNM. In the next section, we shall show quantitative 
results to confirm this guess.

\subsubsection{Buonanno-Chen-Vallisneri template performances}

In Table~\ref{Tab2} we see that the BCV and BCV$_{\rm impr}$ families give almost the same FFs for 
relatively low-mass binary systems 
($M=30,\, 40M_\odot$), while the BCV$_{\rm impr}$ family gives slightly better FFs for higher 
mass binary systems ($M=60,\, 100M_\odot$). For higher-mass binary systems, we find that the $\alpha$ 
parameter takes negative values with reasonable magnitude. This is because the 
amplitude of the NR waveforms in the frequency domain deviates from the $f^{-7/6}$ power 
law only near the merger, which lasts for about {\it one} GW cycle. This merger cycle is important 
only when the total mass of the binary is high enough (see Fig.~\ref{fig:SNRdist}). [See 
also Ref.~\cite{aei} where similar tests have been done.]

The BCV and BCV$_{\rm impr}$ template families give FFs nearly as 
high as those given by the SPA$_c^{\mathcal{Y}}$(4) family, but the latter has 
the advantage of being parametrized directly in terms of the physical binary 
parameters, and it gives fairly small systematic errors.

\subsection{Frequency-domain templates for inspiral, merger and ring-down}
\label{sec4.3}

In this section, we extend our comparisons between the SPA$_c$ 
families and NR waveforms to higher total-mass binary systems ($40M_\odot$ to $120M_\odot$) and 
to unequal-mass binary systems with mass-ratios $m_2/m_1 = 1.5$ and $2$. 
The numerical simulations for unequal-mass binary systems are from the Goddard group. 
They last for $\simeq 373M$ and $\simeq 430M$, respectively, and the NR waveforms have $\simeq 4$ 
cycles before the merger. 

\begin{figure}
\includegraphics[width=0.5\textwidth]{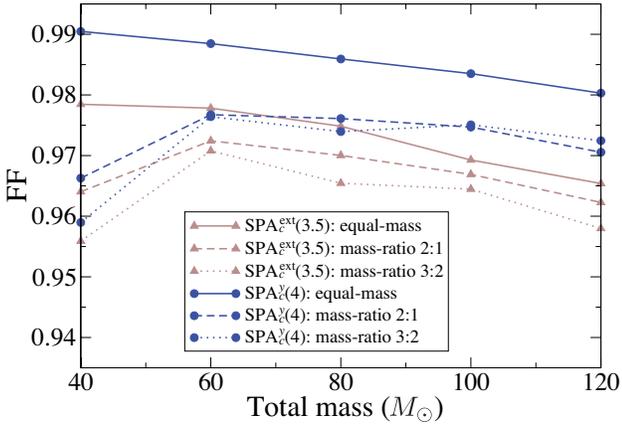}
\caption{FFs as functions of the total binary mass. The FFs are computed between either the 
SPA$_c^{\rm ext}$(3.5) or the SPA$_c^{\mathcal{Y}}$(4) templates and the NR waveforms for equal-mass and 
unequal-mass binary systems.
\label{fig:SPA1}}
\end{figure}

\begin{figure}
\includegraphics[width=0.5\textwidth]{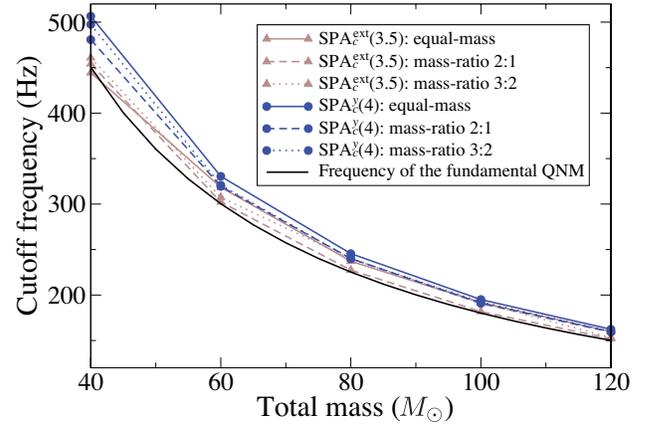}
\caption{Cutoff frequencies as functions of the total binary mass. We show the best-match $f_{\rm cut}$  
for SPA$_c^{\rm ext}$(3.5) and SPA$_c^{\mathcal{Y}}$(4) templates of Fig.~\ref{fig:SPA1}. The solid black curve 
is the fundamental QNM frequency $\omega_{220}/2\pi$. The frequencies are in units of Hz.
\label{fig:SPA2}}
\end{figure}

In Figs.~\ref{fig:SPA1} and \ref{fig:SPA2} we show the FFs for SPA$_c^{\rm ext}$(3.5) 
and SPA$_c^{\mathcal{Y}}$(4) templates, and the values of $f_{\rm cut}$ that achieved 
these FFs~\footnote{Note that because of the short NR waveforms for unequal-mass binary systems, 
we need to search over the starting frequency of templates with a coarse grid, and this causes some 
oscillations in our results. The oscillations are artificial and will be smoothed out in real searches. For instance, 
the drop of FFs at $40M_\odot$ for unequal-mass binary systems happens because the NR waveforms 
are too short and begin right in the most sensitive band of LIGO.}. 
For all mass combinations (except for $M=40M_\odot$ for artificial reasons) the FFs of SPA$_c$(3.5) 
templates are higher than 0.96, and the FFs of SPA$_c^{\mathcal{Y}}$(4) templates are higher 
than 0.97, confirming that both families of templates can be used 
to search for GWs from coalescing binary systems with equal-masses as large 
as $120 M_\odot$ and mass ratios $m_2/m_1 =2$ and 1.5. 
Figure~\ref{fig:SPA2} shows that all the $f_{\rm cut}$ 
values from our searches are within $10\%$ larger than the frequency of 
the fundamental QNM $\omega_{220}$ of an equal-mass 
binary. We have checked that if we fix $f_{\rm cut} = 1.07 \omega_{220}/2\pi$, 
the FFs drop by less than 1\%.

In Fig.~\ref{fig:qnmfreqeta}, we show the same information as in Fig.~\ref{fig:qnmfreq}, 
except that here we draw $\omega_{lnm}$ and $\tau_{lnm}$ as functions of the mass-ratio $\eta$ 
of a nonspinning binary. We compute the spin of the final BH in units of the mass 
of the final BH using the quadratic fit given by Eq. (3.17a) of Ref.~\cite{Jena07long}:
\beq
\frac{a_f}{M_f} \simeq 3.352\eta-2.461\eta^2\,.
\eeq
As Fig.~\ref{fig:qnmfreqeta} shows, $\omega_{220}$ does not change much, confirming the 
insensitivity of the $f_{\rm cut}$ on $\eta$.

However, in real searches we might request that the template family have some deviations 
from the waveforms predicted by NR. For example, a conservative template bank might 
cover a region of $f_{\rm cut}$ ranging from the Schwarzschild innermost stable 
circular orbit (ISCO) frequency, or the innermost circular orbit (ICO) frequency 
determined by the 3PN conservative dynamics, up to a value slightly higher than the 
frequency of the fundamental QNM. The number of templates required to cover the 
$f_{\rm cut}$ dimension depends on the binary masses. We find that to cover the $f_{\rm cut}$ 
dimension from the 3PN ICO frequency to the fundamental QNM frequency with an SPA$_c^{\rm ext}$(3.5) 
template bank, imposing a mismatch $<0.03$ between neighboring templates, we need only 
two  ($\sim$20) templates if $M=30M_\odot$ ($M=100M_\odot$) and $\eta=0.25$. 
In the latter case, the match between templates is more 
sensitive to $f_{\rm cut}$ since most signal power comes from the last two 
cycles, sweeping through a large frequency range, right in 
LIGO's most sensitive band. The number of templates directly affects
the computational power needed, and the false-alarm rate. Further investigations are 
needed in order to determine the most efficient way to search over the 
$f_{\rm cut}$ dimension.

\begin{figure}
\includegraphics[width=0.5\textwidth]{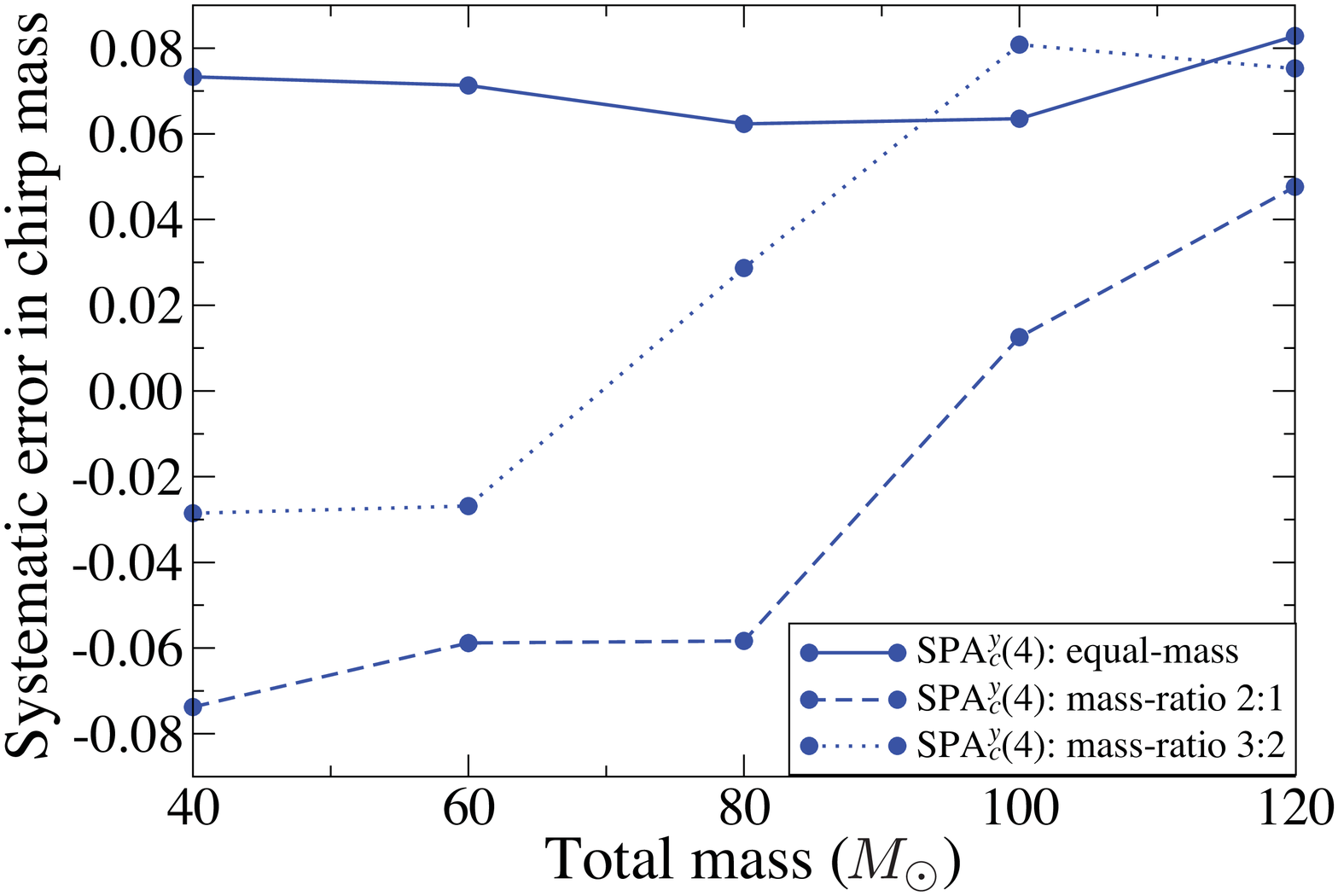}
\caption{Systematic errors of the chirp mass as functions of the total binary mass 
when SPA$_c^{\mathcal{Y}}$(4) templates are used. We show errors of the chirp masses that 
optimize the FFs of Fig.~\ref{fig:SPA1}. 
\label{fig:SPA3}}
\end{figure}

\begin{figure}
\vspace{1.5cm}
\includegraphics[width=0.5\textwidth]{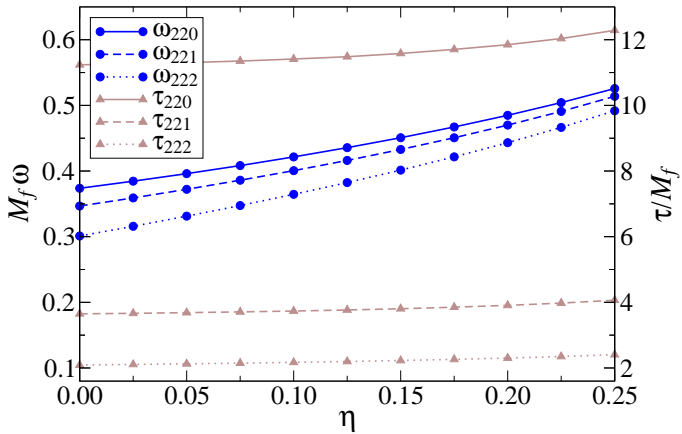}
\caption{Frequencies and decay times of the least damped QNM 220, and two overtones 221 and 222. 
The scales of the frequency and the decay time are listed on the left and right sides of the plot, respectively.
\label{fig:qnmfreqeta}}
\vspace{0.5cm}
\end{figure}

For the purpose of parameter estimation, Fig.~\ref{fig:SPA3} shows that the SPA$_c^{\mathcal{Y}}$(4) templates 
are rather faithful, giving reasonable estimates of the chirp mass: systematic errors less than about $8\%$ in 
absolute value for binary systems with $M=40M_\odot$ up to $M=120M_\odot$. A difference 
of $\simeq 8\%$ may seem large, but the SPA$_c^{\mathcal{Y}}$(4) templates are not exactly 
physical, and more importantly, for large-mass binary systems, most of 
the information on the chirp mass comes {\it only} from the 
last cycle of inspiral. We notice that when the total binary mass is higher than $120M_\odot$, 
the FFs are relatively high (from 0.93 to 0.97), and the estimates of the chirp 
mass are still good (within 10\%). However, for binary systems with such high total 
masses, the ring-down waveform dominates the SNR, and the SPA$_c^{\mathcal{Y}}$(4) template 
family becomes purely phenomenological. A direct ring-down search might be more efficient.


All results for unequal-mass binary systems are obtained using the $C_{22}$ 
component of $\Psi_4$~\cite{BCP}, which is the leading order quadrupole term 
contributing to the GW radiation. For unequal-mass binary systems, higher-order 
multipoles can also be important, and we need to test the performance of 
the template family directly using $\Psi_4$. For $\Psi_4$ extracted in the direction perpendicular 
to the binary orbit, we verified that higher-order multipoles do not appreciably change the FFs.

A natural way of improving the SPA$_c$ models would be to replace 
the discontinuous frequency cut with a linear combination of Lorentzians. We show here a first attempt at doing so. 
The Lorentzian $\mathcal{L}$ is obtained as a Fourier transform of a damped 
sinusoid, e.g., for the fundamental QNM we have
\bea
&& \int_{- \infty}^{\infty} e^{i 2 \pi f t} \, \left ( e^{ \pm i \omega_{220} t - |t|/\tau_{220}} \right )\, dt = \nonumber \\
&&
\frac{2/\tau_{220}}{1/\tau_{220}^2+ (2 \pi f \pm \omega_{220})^2} \equiv 2 {\cal L}_{220}^\pm(f)
\eea
and the (inverse) Fourier transform of Eq.~(\ref{qnmw}) reads
\beq
\sum_n\tilde{h}_{\rm QNM}(f) = A_n\,\left [ {\cal L}_{22n}^+(f)\,e^{i \phi_n} + {\cal L}_{22n}^-(f)\,e^{-i \phi_n} \right ]\,.
\eeq
Restricting to positive frequencies we only keep the ${\cal L}_{22n}^-(f)$ terms. 
In the frequency domain we attach the fundamental mode continuously to the SPA$_c^{\mathcal{Y}}$(4) 
waveform at the ring-down frequency $\omega_{220}$ by tuning the amplitude and 
phase $A_0$ and $\phi_0$. We denote this model SPA$_{\mathcal{L}1}$ (note that 
we also need to introduce the mass-parameter of the final BH as a 
scale for $\omega_{220}$ and $\tau_{220}$). Similarly, we 
define the SPA$_{\mathcal{L}3}$ model where all three QNMs are combined. 
With the three amplitudes and phases as parameters, this model is similar 
to the spin-BCV template family~\cite{bcv2} and we can optimize automatically 
over the 6 parameters. As an example, we compute the 
FFs between the SPA$_{\mathcal{L}1}(4)$ or SPA$_{\mathcal{L}3}(4)$ and 
the NR waveform of an equal-mass $M=100M_\odot$ binary. Using the LIGO PSD, 
we obtain 0.9703 and 0.9817, respectively. Those FFs are comparable 
to the FFs obtained with the simpler SPA$_c$ model, shown in Fig.~\ref{fig:SPA1}. 
It is known that adding more parameters increases the FFs but also increases
the false-alarm probability. By further 
investigation and comparison with NR waveforms our goal is to 
express the phase and amplitude parameters of the Lorentzian in terms of 
the physical binary parameters, relating them 
to the amplitudes and phases of the QNMs and the physics of the 
merger. Those parameters are somewhat similar to the $\epsilon$-parameters 
introduced above for the EOB model when modeling the merger and ring-down 
phases.

We wish to emphasize that the results we presented in this section are preliminary, in the sense that we considered 
only a few mass combinations and the NR waveforms of unequal-mass binary systems are quite short. 
Nevertheless, these results are interesting enough to propose a systematic study of the efficiency 
of these template families through Monte Carlo simulations in real data.

\section{Conclusions}
\label{sec5}

In this paper we compared NR and analytic waveforms emitted by nonspinning binary systems,  
trying to understand the performance of PN template families developed during the last 
ten years and currently used for the search for GWs 
with ground-based detectors, suggesting possible improvements. 

We first computed ${\rm FF}_0$s (maximized only on time and phase) between PN template 
families which best match NR waveforms~\cite{BCP,Goddshort}, 
i.e., Tpn(3), Tpn(3.5) and Epn(3.5). We showed how the drop in ${\rm FF}_0$s is not simply 
determined by the accumulated phase difference between waveforms, but also depends on 
the detector's PSD and the binary mass. Thus, waveforms which differ even by 
one GW cycle can have ${\rm FF}_0 \sim 0.97$, depending on the binary masses 
(see Fig.~\ref{fig:dphovp}).

We then showed that the NR waveforms from the high-resolution and medium-resolution simulations 
of the Goddard group are close to each other (FF$_0$ around 0.99, see Fig.~\ref{fig:NRff}).  
We also estimated that the FF$_0$ between high-resolution and exact NR waveforms 
is even higher, based on the numerical convergence rates of the Goddard simulations. 

Second, by stitching PN waveforms to NR waveforms we built hybrid waveforms, and computed 
${\rm FF}_0$s (maximized only on time and phase) between hybrid waveforms constructed with different 
PN models, notably Tpn(3), Tpn(3.5) and Epn(3.5) models.
We found that for LIGO's detectors and equal-mass binary systems with total mass 
$M>30M_\odot$, the last 11 GW cycles plus merger and ring-down phases contribute 
$>90\%$ of the signal power. This information can be used to set the 
length of NR simulations. 

The ${\rm FF}_0$s between hybrid waveforms are summarized in Figs.~\ref{fig:mmplot}, \ref{fig:mmplotAV}. 
We found that for LIGO's detectors and binary systems with total mass higher than $10M_\odot$, 
the current NR simulations for equal-mass binary systems are long enough to reduce the 
differences between hybrid waveforms built with the PN models 
Tpn(3), Tpn(3.5) and Epn(3.5) to the level of $<3\%$ mismatch. 
For GW detectors with broader bandwidth like advanced LIGO and VIRGO, longer 
NR simulations will be needed if the total binary masses $M<10M_\odot$. 
With the current available length of numerical simulations, it is hard to 
estimate from the FFs between hybrid waveforms how long the simulations should be. 
Nevertheless, from our study of the distribution of signal power, we estimate 
that for $M<10M_\odot$ binary systems, at least $\sim80$ NR inspiraling cycles before 
merger are needed.

Finally, we evaluated FFs (maximized on binary masses, initial time 
and phase) between full NR (or hybrid waveforms, depending on the 
total binary mass) and several time and frequency domain PN 
template families. For time-domain PN templates and binary 
masses $ 10 M_\odot < M < 20 M_\odot$, for which the merger/ring-down 
phases do not contribute significantly to the total detector 
signal power, we confirm results obtained 
in Refs.~\cite{BCP,Goddshort}, notably that Tpn(3.5) and Epn(3.5) models 
have high FFs with good parameter estimation, i.e., they are faithful. We found that 
the frequency-domain SPA family has high FFs only for binary systems with 
$M<20M_\odot$, for which most of the signal power comes from the 
early stages of inspiral. Furthermore, we found that it is possible 
to improve the SPA family by either extending 
it to unphysical regions of the parameter space (as done with BCV templates) 
or by introducing an {\it ad hoc} 4PN-order constant coefficient in the phase. 
Both modified SPA families achieve high FFs for high-mass 
binary systems with total masses $30M_\odot<M<120M_\odot$.

\begin{figure}
\includegraphics[width=0.5\textwidth,clip=true]{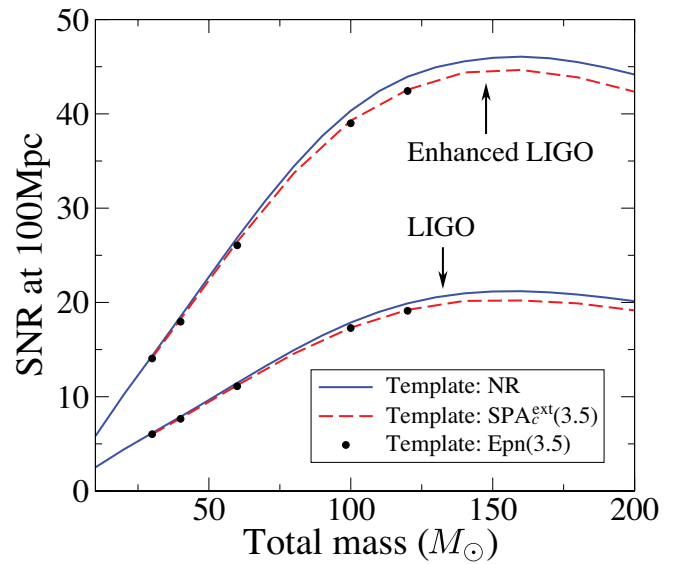}
\caption{The sky-average SNR for LIGO and Enhanced or mid LIGO detector versus total mass for an equal-mass 
binary at 100Mpc.
\label{fig:SPASNR}}
\end{figure}

For time-domain PN templates and binary 
masses $ M \,\gaq\, 30 M_\odot$, we found that if a superposition of ring-down modes 
is attached to the inspiral waveform, as naturally done in the 
EOB model, the FFs can increase from $\sim0.8$ to $>0.9$. 
We tested the current Epn(3.5) template family obtained by 
attaching to the inspiral waveform three QNMs~\cite{BCP} around the EOB 
light-ring. In order to properly take into account the energy and angular-momentum 
released during the merger/ring-down phases we introduced~\cite{BCP} two physical 
parameters, $\epsilon_M$ and $\epsilon_J$, whose dependence on 
the binary masses and spins will be determined by future comparisons 
between EOB and NR waveforms computed for different mass ratios and spins. 
We found high FFs $\gaq 0.96$. Due to small differences between EOB and NR 
waveforms during the final cycles of the evolution, the best-matches 
are reached at the cost of large  systematic error in the merger--ring-down binary parameters. 
Thus, the Epn(3.5) template family can be used for detection, but 
for parameter estimation it needs to be improved when matching to the 
ring-down, and also during the inspiral phase. The refinements can be achieved 
(i) by introducing deviations from circular motion, (ii) adding higher-order PN 
terms in the EOB dynamics, (iii) using 
in the EOB radiation-reaction equations a GW energy flux closer the the NR flux,
(iv) designing a better match to ring-down modes, 
etc.. The goal would be to achieve dephasing 
between EOB and NR waveforms of less than a few percent in the comparable-mass case, as 
obtained in Ref.~\cite{DNT} in the extreme mass-ratio limit.
Indeed, with more accurate numerical simulations, especially 
those using spectral methods~\cite{caltech-cornell}, it will be possible 
to {\it improve} the inspiraling templates by introducing higher-order 
PN terms in the analytic waveforms computed by direct comparison with 
NR waveforms. 

Frequency-domain PN templates with an appropriate cutoff frequency $f_{\rm cut}$ 
provide high FFs ($>0.97$), even for large masses. This is due to oscillating 
tails (Gibbs phenomenon) produced when cutting the signal in the frequency domain. 
We tested the SPA$_c^{\rm ext}$(3.5) and the SPA$_c^{\mathcal{Y}}$(4) template families for total masses up to $120M_\odot$, 
and three mass ratios $m_2/m_1 =1, 1.5$, and 2. We always get FFs $>0.96$, even   
when using a fixed cutoff frequency, $f_{\rm cut} = 1.07\omega_{220}/2\pi$. 
Because of its high efficiency, faithfulness, i.e., low  systematic error in parameter estimation, and simple 
implementation, the SPA$_c^{\mathcal{Y}}$(4) template family (or variants of it which 
include Lorentzians) is, together with the 
EOB model, a good candidate for searching coherently for GWs from binary systems 
with total masses up to $120M_\odot$. 

In Fig.~\ref{fig:SPASNR}, we show the sky averaged SNRs of a single LIGO and Enhanced or mid LIGO \cite{eligo} detector, 
for an equal-mass binary at 100Mpc. The SNR peaks at the total binary mass 
$M \simeq 150M_\odot$ and shows the importance of pushing current searches 
for coalescing binary systems to $M>100M_\odot$. In the mass range $30M_\odot<M<120M_\odot$, 
the SNR drops only slightly if we filter the GW signal with SPA$_c^{\rm ext}$(3.5) or 
Epn(3.5) instead of using NR waveforms. The difference between Epn(3.5) and 
SPA$_c^{\rm ext}$(3.5) is almost indistinguishable. When 
$M>120M_\odot$, although the SPA$_c^{\rm ext}$(3.5) and Epn(3.5) template families give 
fairly good SNRs, it is maybe not a good choice to use them as the 
number of cycles reduces to a few. The key problem in detecting such GWs is how to veto triggers 
from non-Gaussian, nonstationary noise, instead of matching the effectively short signal. 
This is a general problem in searches for short signals in ground-based detectors.

\acknowledgments
A.B. and Y.P. acknowledge support from NSF grant PHY-0603762, and A.B. also 
from the Alfred Sloan Foundation. 
The work at Goddard was supported in part by NASA grants
O5-BEFS-05-0044 and 06-BEFS06-19. B.K. was supported by the NASA Postdoctoral 
Program at the Oak Ridge Associated Universities.
S.T.M. was supported in part by the Leon A. Herreid Graduate Fellowship.
Some of the comparisons with PN and EOB models were obtained building on 
Mathematica codes developed in Refs.~\cite{BD2,bcv1,bcv2,pbcv}

\appendix

\section{Comment on waveforms obtained from the  energy-balance equation}
\label{appA}

\begin{figure*}
\includegraphics[width=0.5\textwidth]{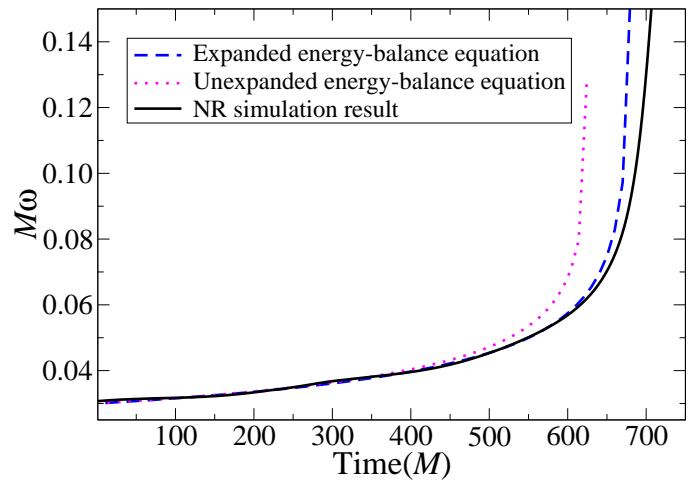}
\caption{Orbital frequency evolution. The dotted and dashed 
curves are calculated from the unexpanded and expanded 
energy-balance equations. The continuous curve refers to 
the really long Goddard NR simulation. 
\label{fig:omegas}}
\end{figure*}

In adiabatic PN models, like the Tpn model used in this paper, waveforms are 
computed under the assumption that the binary evolves through an adiabatic sequence 
of quasi-circular orbits. More specifically, one sets $\dot{r}=0$ and computes the 
orbital frequency $\omega$ from 
the energy-balance equation $dE(\omega)/dt=\mathcal{F}(\omega)$, where $E(\omega)$ 
is the total energy of the binary system and $\mathcal{F}(\omega)$ 
is the GW energy flux. Both  $E(\omega)$ and $\mathcal{F}(\omega)$ are 
computed for circular orbits and expressed as a Taylor expansion in  
$\omega$. The adiabatic evolution ends {\it in principle} 
at the innermost circular orbit (ICO)~\cite{ICO}, or minimum energy 
circular orbit (MECO)~\cite{bcv2}, where $(dE/d \omega)=0$.

By rewriting the energy-balance equation, $\omega(t)$ can be 
integrated directly as
\beq\label{eb}
\dot{\omega}(t)=\frac{\mathcal{F}(\omega)}{dE(\omega)/d\omega}.
\eeq
The RHS of Eq.~\eqref{eb} can be expressed as an expansion 
in powers of $\omega$. The expanded version is widely used in generating 
adiabatic PN waveforms~\cite{DIS01,bcv1,bcv2,pbcv}, it is 
used to generate the so-called 
Tpn template family. It turns out that Tpn(3) and Tpn(3.5) are 
quite close to the NR inspiraling waveforms~\cite{BCP,Goddshort}. 
We wonder whether using the energy-balance in the form of Eq.~(\ref{eb}), 
i.e., without expanding it, might give PN waveforms closer to or farther 
from NR waveforms. In principle the adiabatic 
sequence of circular orbits described by Eq.~(\ref{eb}) ends 
at the ICO, so the adiabatic model should work better 
until the ICO and start deviating (with $\omega$ going to 
infinity) from the exact result beyond it. 

In Fig.~\ref{fig:omegas} we show the NR orbital frequency $\omega(t)$  
together with the PN orbital frequency obtained by solving the unexpanded  
and expanded form of the energy-balance equation. The frequency evolution 
in these two cases is rather different, with  
the orbital-frequency computed from the expanded energy-balance 
equation agreeing much better with the NR one. 
When many, extremely accurate, GW cycles from NR will be available, 
it will be worthwhile to check whether this result is still true.

\end{document}